\documentclass[aip,jcp, numerical, reprint]{revtex4-1}
\usepackage{graphicx}
\graphicspath{{images/}}
\usepackage[table,xcdraw]{xcolor}
\usepackage{amsmath,amsfonts}
\usepackage{mathrsfs}
\usepackage[version=3]{mhchem}
\usepackage{mathptmx, bm}
\usepackage{float}
\usepackage{enumerate}
\usepackage{enumitem}
\usepackage{booktabs}
\usepackage{makecell}
\usepackage{array} 
\DeclareMathSizes{10}{9}{7}{5} 

\usepackage{pifont}
\newcommand{\cmark}{\ding{51}}%
\newcommand{\xmark}{\ding{55}}%

\begin{document}
\newcommand \e[1]{{\textbf{#1}}}
\newcommand \bo[1]{{\bf{{#1}}}}
\newcommand \bra[1] {\left\langle {#1} \right\vert}
\newcommand \ket[1] {\left\vert {#1} \right\rangle }
\newcommand \braket[2] { \left\langle {#1}  \middle\vert{#2} \right\rangle}
\newcommand \braketthree[3] { \left\langle {#1}  \middle\vert{#2} \middle\vert{#3} \right\rangle}
\newcommand{\RNum}[1]{\uppercase\expandafter{\romannumeral #1\relax}}


\title{On the electronic path integral normal modes of the Meyer--Miller--Stock--Thoss representation of nonadiabatic dynamics}
\author{Lauren E. Cook}
\author{Timothy J. H. Hele}
 \email{\href{mailto:t.hele@ucl.ac.uk}{t.hele@ucl.ac.uk}}
 \affiliation{Department of Chemistry, University College London, Christopher Ingold Building, London WC1H 0AJ, United Kingdom}

\date{\today}

\begin{abstract}
Accurate and efficient simulation of nonadiabatic dynamics is highly desirable for understanding charge and energy transfer in complex systems. A key criterion for obtaining an accurate method is conservation of the Quantum Boltzmann Distribution (QBD).
For a single surface, Matsubara dynamics is known to conserve the QBD, as a consequence of truncating the dynamics in the higher normal modes of the imaginary-time path integral. Recently, a nonadiabatic Matsubara (NA-Mats) dynamics has been proposed (J. Chem. Phys., 2021, 154, 124124) which truncates in the normal modes of the nuclear variables but not in the electronic variables, which are described with the Meyer--Miller--Stock--Thoss (MMST) representation. Surprisingly, this NA-Mats method does not appear to conserve the QBD for a general system. 
This poses the question of the effect of truncating the higher path integral normal modes of the \emph{electronic} variables in the MMST representation. 
In this article, we present what we believe is the first study of electronic normal modes of the MMST representation. We find that observables are not usually a function of a finite number of normal modes and that the higher normal modes are not constrained by the distribution, unlike in conventional nuclear normal modes. 
Furthermore, truncating the dynamics in MMST normal modes leads to inaccurate correlation functions and while the QBD appears conserved for an ensemble of trajectories, it is not for a single trajectory.
Overall, this suggests that MMST path integral normal modes are not optimal for obtaining an accurate, QBD conserving nonadiabatic dynamics method.  
\end{abstract}

\pacs{}

\maketitle 


\section{Introduction}
Many of the interesting and complex phenomena observed from charge and energy-transfer mechanisms occur via nonadiabatic dynamics, where the electronic and nuclear degrees of freedom are coupled. Examples include photoisomerism, ultrafast relaxation via intersystem crossing in DNA or charge carriers in solar cell materials, and electron transfers in various photosynthetic systems and catalytic materials.\cite{polliConicalIntersectionDynamics2010, wanFemtosecondDynamicsDNAmediated1999, hammes-schifferTheoryCoupledElectron2010, marcusElectronTransfersChemistry1985, zhuChargeCarriersHybrid2015,chengDynamicsLightHarvesting2009, domckeRoleConicalIntersections2012} As these processes often occur on very short timescales,\cite{mukherjee_assessing_2025} it can be experimentally challenging to understand the dynamics. 

Hence, many theoretical methods have been developed to simulate nonadiabatic dynamics.\cite{Althorpe2016, nelsonNonadiabaticExcitedStateMolecular2020} These range from full quantum methods such as Multi-Configurational Time-Dependent Hartree-Fock (MCTDH) and wavepacket propagation,\cite{Beck2000,vanhaeftenPropagatingMultidimensionalDensity2023,wangMultilayerFormulationMulticonfiguration2003, shinMultipleTimeScale1996} to more approximate semiclassical methods including initial-value representations (IVR),\cite{Stock1997, Miller1970, Wang1998, Sun1998, Sun1998b, Church2018, Ananth2007} nonadiabatic ring-polymer molecular dynamics (RPMD) and their derivatives.\cite{Richardson2013, Ananth2013, Hele2011, Hele2015, Cao1993, Cao1994, Cao1994a, Cao1994b, Rossi2014} Another useful class of methods are mixed quantum-classical methods, including surface hopping, Ehrenfest dynamics, and mapping-based methods.\cite{Kapral2016, Tully1971, tullyPerspectiveNonadiabaticDynamics2012, Shakib2017, Shalashilin2011, Zimmermann2014, Meyer1979,Stock1997, Stock2005, Runeson2019, MASH} These methods often have a good balance between computational cost and accuracy. 

In this article, we will focus on the Meyer-Miller-Stock-Thoss (MMST) mapping method where the quantum degrees of freedom (DoF) are mapped onto discrete classical DoF that can be propagated. This mapping is well-established and maps onto electronic position and momenta, sometimes expressed in action-angle variables, where the former are intuitive to propagate with classical mechanics.\cite{Meyer1979, Stock1997} This is not the only choice of well-established mapping method as there have been many recent developments in spin-mapping,\cite{Runeson2019, Runeson2020, Runeson2021} causing a surge in popularity of these methods. These developments include the Mapping Approach to Surface Hopping (MASH) and extension to multiple electronic states.\cite{Amati2023, Mannouch2020, MASH, richardsonNonadiabaticDynamicsMapping2025, runeson_exciton_2024, lawrence_size-consistent_2024, geutherTimeReversibleImplementationMASH2025} Accurate algorithms have also been developed to propagate the MMST variables.\cite{Church2018,Cook2023} 

When considering the development of a nonadiabatic dynamics method, it would be highly desirable if the method satisfied the following criteria:
\begin{enumerate}
    \item Conservation the quantum Boltzmann distribution (QBD) 
    \item Reproduction of Rabi oscillations
    \item Classical scaling with the system size 
    \item A derivation from exact quantum dynamics.
\end{enumerate}

While many of the existing methods satisfy many of these criteria, as far as we are aware, none satisfy all of them simultaneously and exactly for a general quantum system.

Conservation of the QBD is a particularly interesting criterion to investigate, as for single-surface dynamics, Matsubara dynamics conserves the QBD.\cite{Willatt2017, Hele2015} Matsubara dynamics achieves this by starting with an (exact quantum) Generalized Kubo Transform,\cite{Hele2013, Hele2013b, Althorpe2013, Hele2014, hele_alternative_2016} constructing the Liouvillian, and then transforming this to path integral normal modes. This transformation results in the observable being a function of a finite number of the lowest normal modes and truncation in the higher normal modes leads to dynamics which is both classical and conserves the QBD. Recent work has proposed a nonadiabatic Matsubara (NA-Mats) method, but to our knowledge, this does not conserve the QBD.\cite{Chowdhury2021, Richardson2017} This is unanticipated as the Matsubara dynamics approach to the nuclear DoF is utilised, so one would expect that it may inherit the same properties. However, the electronic DoF are left unchanged in their MMST variable form. This leaves the question of if additionally following the Matsubara approach with electronic normal modes will lead to QBD conservation for a NA-Mats method.

It would be highly desirable if a nonadiabatic version of Matsubara dynamics could be derived which could be proven to conserve the QBD. Although the dynamics may be too costly to evaluate for large systems (due to a probable phase factor), approximations to this may lead to accurate and inexpensive methods, in the same way that approximations to single-surface Matsubara dynamics lead to RPMD, Centroid Molecular Dynamics (CMD) and Thermostatted (T)-RPMD.\cite{Craig2004,ceriottiEfficientStochasticThermostatting2010, heleCommunicationRelationCentroid2015, Hele2016a,Hele2015, jang_derivation_1999,jang_path_1999, Hele2015c}

In this work, we aim to explore the properties of electronic normal modes to better understand them and to determine whether the conservation of the QBD can be obtained. We will do this by considering an electronic-only system and take normal modes of the MMST representation, as this form is used in the the proposed NA-Mats.\cite{Chowdhury2021} The MMST mapping Hamiltonian is known to satisfy some of our criteria, such as replicating Rabi oscillations and having accurate, classical scaling algorithms for propagation.\cite{Cook2023}

The structure of this article is as follows. In section \ref{backgroundtheory}, we present background theory, including correlation functions and Matsubara dynamics. In section \ref{methodology}, we present the methodology of electronic normal modes of the MMST variables. Algebraically, we determine that truncating in MMST normal modes will not conserve the QBD for a single trajectory. In section \ref{results}, we present the computational results of MMST normal modes for an electronic-only system. We find that all MMST normal modes are required to obtain reasonable dynamics, and that there is no narrowing in the distribution of the higher normal modes. Truncating in normal modes appears to conserve the distribution for an ensemble of trajectories due to an averaging effect. We conclude in section \ref{conclusions}.

\section{Background Theory} \label{backgroundtheory}
\subsection{Single Surface}
For a single potential energy surface (PES), Matsubara dynamics is known to conserve the QBD.\cite{Hele2015} As far as we are aware, for multiple surfaces there does not exist a method that satisfies all our desirable criteria, as listed earlier. A NA-Mats method has been proposed, but it only conserves the distribution in certain limits, which will be discussed later.\cite{Chowdhury2021} Here, we will discuss how the Matsubara dynamics conserves the QBD for a single surface, which has also been used for the nuclear variables in the NA-Mats method (2021).\cite{Chowdhury2021}

The quantum Hamiltonian for a single-surface is, 
\begin{align} \label{quntum-ham}
    \hat{H} = \frac{\hat{\bo{P}}^2}{2m} +V(\hat{\bo{R}}) \text{,}
\end{align}
where $\hat{R}$ and $\hat{P}$ are the nuclear position and momentum operators respectfully, and $V(\hat{\bo{R}})$ is the potential energy. 

Quantum dynamics methods are commonly used to calculate correlation functions (CF) as these can be used to extract properties of the system that can be compared to experiment. These can take the standard (unsymmetrized) form, 
\begin{align}
    \label{General-CF}
    C_{AB}= \frac{1}{Z}\mathrm{Tr}[e^{-\beta \hat{H}}\hat{A}(0)\hat{B}(t)]\text{,}
\end{align}
where $e^{-\beta \hat{H}}$ is the QBD, $\hat{A}$ and $\hat{B}$ are operators where $\hat{B}(t) = e^{i\hat{H}t/\hbar}\hat{B}(0) e^{-i\hat{H}t/\hbar}$ and $Z =\mathrm{Tr}[e^{-\beta \hat{H}}]$ is the partition function. The related Kubo-transformed (KT) CF has a higher degree of symmetry than the standard CF in Eqn.~\eqref{General-CF},\cite{kuboFluctuationdissipationTheorem1966, Craig2004, Willatt2017} 
\begin{align} \label{kubotransformed}
    C_{AB}(t) = \frac{1}{Z \beta} \int^{\beta}_{0} d\lambda \text{Tr} \left[ e^{-(\beta-\lambda)\hat{H}} \hat{A} e^{-\lambda\hat{H}} e^{i\hat{H}t/\hbar} \hat{B} e^{-i\hat{H}t/\hbar} \right]\text{,}
\end{align}
which satisfies detailed balance and reality.\cite{Craig2004, Willatt2017} In addition, $C_{AB}(t) = C_{AB}(-t)^*$ (time-reversal symmetry) will be satisfied for both the KT-CF and standard CF when the operators are functions of only position.\cite{Willatt2017, Craig2004} One can obtain the standard CF from the KT-CF (and vice-versa) through a Fourier Transform, which makes it advantageous to calculate the KT-CF due to this higher degree of symmetry.\cite{Craig2004}

When evaluating a CF in ring-polymer co-ordinates, a discretised imaginary-time path integral version of the KT-CF is used to describe the evolution of the path integral as a whole. Each bead is then a slice of the path integral in imaginary time, and cyclic evaluation of this generalised KT-CF around the ring-polymer is an approximation to the whole path integral. The generalised KT-CF is,\cite{Hele2015, Hele2013, Hele2013b, Althorpe2013} 
\begin{align}
    \label{generalised-KT-cf}
    C^{[N]}_{AB}(t) 
    &=  \frac{1}{Z} \prod^{N}_{j} \iint d{\bf{R}}_{j} d{\boldsymbol{\mu}}_{j} \nonumber \\ & \quad \times \biggl\langle {R_{j-1} - \mu_{j-1}/2} \bigg\vert {\frac{1}{2}(\hat{A}e^{-\beta_N\hat{H}} + e^{-\beta_N\hat{H}}\hat{A})} \bigg\vert  R_j + {\mu}_j/2 \biggr\rangle \nonumber \\ & \quad \times \braketthree{{R}_j + {\mu}_j/2} {e^{i\hat{H}t/\hbar} \hat{B} e^{-i\hat{H}t/\hbar}}{R_j - \mu_j/2} \text{,}
\end{align}
where the operators act once in the path integral such that $\hat{A} = (\sum_j^N \hat{A}_j)/N$ and likewise for $\hat{B}$, the indices are cyclic ($R_N =R_0$) and $\beta_N= \beta/N$. The generalised KT-CF is then equivalent to the KT-CF in the $N \to \infty$ limit when the operators can be written as a linear sum of operators that act once in the path integral, as is defined above.\cite{Hele2013, Willatt2017, Hele2015} For example, the flux and side operators utilised in rate theory are non-linear under this definition.\cite{Althorpe2013, Hele2013, Hele2013b, Althorpe2013, Hele2014} Matsubara dynamics approximates the generalised KT-CF through truncation of the Liouvillian which is discussed in further detail later.\cite{Hele2015}


The Wigner transform can be utilised to obtain,\cite{Hele2016, Wigner1932,hilleryDistributionFunctionsPhysics1984} 
\begin{align} \label{GKT_wigner-nuc}
    C^{[N]}_{AB}(t) &= \frac{1}{Z(2\pi\hbar)^{N} }\iint d{\bf{R}} d{\bf{P}} [e^{-\beta_N \hat{H}}\hat{A}]_{\overline{W}} \times [\hat{B}(t)]_W \text{,}
\end{align}
where the real-time and imaginary-time terms are defined as, 
\begin{subequations} \label{wigner-rt-it-nuc}
\begin{align}
    &[\hat{B}(t)]_W = \prod^{N}_{j} \int d{\mu}'_{j} e^{i{\boldsymbol{\mu}}'_j {P}_j/\hbar}  \braketthree{{R}_j + \mu'_j/2} { \hat{B}(t) }{R_j - \mu'_j/2} \text{,}\\ 
    &[e^{-\beta_N \hat{H}}\hat{A}]_{\overline{W}} = \prod^{N}_{j} \int d{\mu}_{j}  e^{i{\mu}_j {P_j/\hbar}} \left. \biggl\langle {R_{j-1} - \mu_{j-1}/2} \bigg\vert \mathcal{A} \middle\vert  R_j + {\mu}_j/2 \right\rangle \text{,}
\end{align}
\end{subequations}
where $\mathcal{A} = (\hat{A}e^{-\beta_N\hat{H}} + e^{-\beta_N\hat{H}}\hat{A})/2$ is the symmetric operator, $\hat{B}(t) = e^{i\hat{H}t/\hbar} \hat{B} e^{-i\hat{H}t/\hbar}$ is the time-evolved operator and, $\int d\bo{R} = \prod_j^N \int d \bo{R}_j$ and likewise for $P$ and $\mu$.\cite{Hele2015, Hele2016} The form of $[\cdot]_{\overline{W}}$ and $[\cdot]_W$ are different, where the latter is a simple bead-average of the Wigner transform and the former is more complex. 

Matsubara dynamics uses the normal modes of a free ring-polymer which are well defined as,  
\begin{align}
    \check{R}_{k} = \sum_{j=1}^N T_{jk}R_{j} \text{,}
\end{align}
where $j$ is the bead index and $k$ is the mode index. The transformation matrix, $\bf{T}$, is chosen to diagonalize the free ring-polymer spring constant matrix within the ring-polymer Hamiltonian,\cite{ceriottiEfficientStochasticThermostatting2010} and for an even number of beads is, 
\begin{align}
    \label{transformation}
    T_{jk} = \begin{cases}
        \sqrt{\frac{1}{N}} \quad & k=0 \\
        \sqrt{\frac{2}{N}} \cos(2\pi jk/N) \quad &1 \leq k \leq N/2-1 \\
        \sqrt{\frac{1}{N}}(-1)^j \quad & k=N/2 \\
        \sqrt{\frac{2}{N}} \sin(2\pi jk/N) \quad & N/2-1 \leq k \leq N-1 \text{,} 
    \end{cases} 
\end{align}
which is equivalent to the half-complex Fourier transform. We note that the Matsubara Hamiltonian does not explicitly have springs connecting adjacent beads. 
The normal modes have frequencies, 
\begin{align}
\label{normalmodefre}
\omega_k = \frac{2}{\beta_N \hbar}\sin \left(\frac{k \pi}{N}\right) \text{.}
\end{align}
When considering the $M$ lowest normal modes in the limit $N \to \infty$ and $M<< N$, the frequencies of the modes tend to the `Matsubara' frequencies, 
\begin{align}
    \label{mats-freq}
    \tilde{\omega}_k = \lim_{N \to \infty} \omega_k = \frac{2k \pi}{\beta \hbar} \text{,}
\end{align}
and are hence known as the Matsubara modes, 
\begin{align}
    \label{matsubara-modes}
    \tilde{R}_k = \lim_{N \to \infty} \frac{\check{R}_k}{\sqrt{N}} \text{,}
\end{align}
where the additional factor of $\sqrt{N}$ ensures convergence such that $\tilde{R}_0$ is the centroid.\cite{Hele2015, heleCommunicationRelationCentroid2015}

Matsubara modes have been utilized in path integral descriptions of equilibrium properties for many years, and gives rise to an approximate ring-polymer expression for the zero-time value of Eqn.~\eqref{kubotransformed}.\cite{Ceperley1995, chakravartyPathIntegralSimulations1997, chakravartyComparisonEfficiencyFourier1998,freemanMonteCarloMethod1984} This means that many observables can be written as a function of only a finite number of the lowest normal modes of the nuclear position and/or momenta.\cite{Hele2015, hele_alternative_2016} The Matsubara modes have a special property where any superposition of them results in a smooth and differentiable distribution with respect to imaginary time.\cite{Hele2015} The Liouvillian can also be expressed in terms of the normal modes, and can be truncated to be only the lowest normal modes. This is an approximation as the propagation of the higher and lower normal modes is generally coupled. Truncation of the Liouvillian in the higher normal modes surprisingly results in classical dynamics, occurring when the Matsubara modes are decoupled from the non-Matsubara modes.\cite{Hele2015} The (unpropagated) higher normal modes can then be integrated out due to the QBD. This also conserves the QBD, as the remaining lowest path integral modes are a smooth function of imaginary time such that Noether's theorem can be used to prove quantum Boltzmann conservation.\cite{Hele2015, goldsteinClassicalMechanics1980} 

\subsection{Nonadiabatic Systems}
The previously proposed NA-Mats method uses this Matsubara approach for the nuclear DoF, but leaves the electronic DoF unchanged in the MMST mapping representation, and this generally does not conserve the QBD.\cite{Chowdhury2021} We suspect that to achieve conservation of the QBD for a nonadiabatic Matsubara method, a similar approach to Matsubara dynamics for the normal modes of the electronic degrees of freedom may be required. When considering electronic normal modes, there are many metrics that one can choose from for the state populations. For example, the occupancy calculated from Meyer--Miller--Stock--Thoss (MMST) mapping variables,\cite{Meyer1979, Stock2005} spin-mapping or the action-angle approach,\cite{Runeson2019, Runeson2020, Meyer1979} and it is not immediately apparent which metric to use. We ideally would like a metric that satisfies the following,\cite{Hele2015, heleCommunicationRelationCentroid2015}
\begin{enumerate}
    \item Observables are a function of a finite number of the lowest normal modes
    \item The distribution of the higher normal modes narrows (such that they can be integrated out when dynamics is truncated)
    \item Truncating the normal modes results in conservation of the QBD
    \item Truncating the normal modes provides a good approximation to exact (untruncated) dynamics 
\end{enumerate} 
Matsubara dynamics takes normal modes of the nuclear position and momenta, and NA-Mats uses the MMST electronic position and momenta, so the `obvious' choice, and the one we will focus on in this work, is the MMST metric. Although spin-mapping has become increasingly popular in recent years,\cite{Runeson2019, Runeson2020, Runeson2021, richardsonNonadiabaticDynamicsMapping2025, MASH, Amati2023} we leave a detailed analysis of this for future research. 

For a nonadiabatic system, the Hamiltonian for an $F$-level system in the diabatic representation is, 
\begin{equation} \label{ODH}
	\hat{H}= \frac{1}{2}\hat{\bo{P}}^\mathrm{T} \boldsymbol{\mu}^{-1}\hat{\bo{P}} + V_{0}(\hat{\bo{R}})+ \sum_{n,m=1}^{F} \ket{n} \bo{V}_{nm}(\hat{\bo{R}}) \bra{m} \text{,}
\end{equation}
where the diabatic potential matrix in the basis of electronic states, $\ket{n}$, is \e{V}(\e{R}), the state-independent potential matrix is $V_{0}(\bo{R})$, and $\boldsymbol{\mu}$ is a diagonal matrix of nuclear masses. 

The MMST Hamiltonian maps the $F$-level system onto $F$ coupled harmonic oscillators, each of which has an electronic position and momenta, and in the diabatic basis is, 
\begin{align} \label{full-MMST}
    H = \frac{1}{2} \bo{P}^\mathrm{T} \boldsymbol{\mu}^{-1} \bo{P} + V_{0}(\bo{R}) + \frac{1}{2}\left\{ \bo{p}^\textrm{T}\bo{V}(\bo{R})\bo{p} + \bo{q}^\textrm{T}\bo{V}(\bo{R})\bo{q} - \textrm{Tr}[\bo{V}(\bo{R})]\right\} \text{,}
\end{align}
where $\bo{q}$ and $\bo{p}$ are the electronic position and momenta.

 
The exact nonadiabatic Liouvillian generated from a Moyal series in mapping variables is,\cite{Chowdhury2021, Hele2016, Moyal1949}
    \begin{equation}
    \begin{split}
        \label{exact-liouvillian}
        \mathcal{L}_{\mathrm{exact}} = &\sum_{l}^N \frac{P_{l}}{m}\overrightarrow{\frac{\partial}{\partial R_{l}}} -\frac{2}{\hbar} \left[ V_{0}(R_{l}) + V_{e}(R_{l}, {\bf{q}}_{l}, {\bf{p}}_{l})\right] \sin \left( \frac{\hbar}{2} \overleftarrow{\frac{\partial}{\partial R_{l}}} \overrightarrow{\frac{\partial}{\partial P_{l}}}\right) \\ &+  \left[ {\bf{p}}_{l}^{\textrm{T}} {\bf{V}}(R_{l}) \overrightarrow{\nabla}_{{\bf{q}}_{l}} -  {\bf{q}}_{l}^{\textrm{T}} {\bf{V}}(R_{l}) \overrightarrow{\nabla}_{{\bf{p}}_{l}} \right]\frac{1}{\hbar} \cos  \left( \frac{\hbar}{2} \overleftarrow{\frac{\partial}{\partial R_{l}}} \overrightarrow{\frac{\partial}{\partial P_{l}}}\right) \\ &+ \frac{1}{4}  \left[ \overrightarrow{\nabla}_{{\bf{p}}_{l}}^{\textrm{T}} {\bf{V}}(R_{l}) \overrightarrow{\nabla}_{{\bf{p}}_{l}} +  \overrightarrow{\nabla}_{{\bf{q}}_{l}}^{\textrm{T}} {\bf{V}}(R_{l}) \overrightarrow{\nabla}_{{\bf{q}}_{l}} \right] \sin \left( \frac{\hbar}{2} \overleftarrow{\frac{\partial}{\partial R_{l}}} \overrightarrow{\frac{\partial}{\partial P_{l}}}\right) \text{,}
    \end{split}
    \end{equation}
    where $V_{e}(R_{l}, {\bf{q}}_{l}, {\bf{p}}_{l}) = \frac{1}{2}\left( \bo{p}_l^\textrm{T}\bo{V}(R_l)\bo{p}_l + \bo{q}_l^\textrm{T}\bo{V}(R_l)\bo{q}_l - \textrm{Tr}[\bo{V}(R_l)]\right) $ is the electronic part of the Hamiltonian for the $l$-th bead, and the arrow indicates what direction the derivative is taken in. The first line on the RHS corresponds to Ehrenfest-like evolution of nuclear DoF, the second is principally electronic evolution and the third term contains higher-order couplings.\cite{Chowdhury2021} In this article we follow the convention set in the Matsubara paper and by Zwanzig,\cite{Hele2015, Zwanzig2001} and define the Liouvillian without the prefactor of the imaginary unit, $i$, such that it is real.
    
For a system with decoupled electronic and nuclear degrees of freedom, the Liouvillian becomes,\cite{Chowdhury2021} 
\begin{align} \label{decoupled-liouvillian}
    \mathscr{L}_{\textrm{decoupled}} = \sum_{l=1}^{N} \frac{P_{l}}{m} \frac{\overrightarrow{\partial}}{\partial R_{l}} - & V_{0}(R_{l}) \frac{2}{\hbar} \sin\left(\frac{\hbar}{2} \frac{\overleftarrow{\partial}}{\partial R_{l}} \frac{\overrightarrow{\partial}}{\partial P_{l}} \right) \nonumber \\ & + \frac{1}{\hbar} \left[ {\bf{p}}^{\textrm{T}}_{l} {\bf{V}} \overrightarrow{\nabla}_{{\bf{q}}_{l}} -  {\bf{q}}^{\textrm{T}}_{l} {\bf{V}} \overrightarrow{\nabla}_{{\bf{p}}_{l}} \right] \text{,}
\end{align}
which is separable into nuclear and electronic terms. This includes two limits of the system, where there is an electronically adiabatic system and if there is only the electronic system.\cite{Chowdhury2021} This is one of the only cases where the proposed NA-Mats method conserves the QBD.\cite{Chowdhury2021} The second case is when the electronic DoF are linearly coupled to a harmonic bath.\cite{Chowdhury2021}

In this work, we will investigate the use of electronic normal modes in the MMST representation to compute generalised KT-CF and compare to the exact KT-CF, and to check the satisfaction of the metric criteria listed above.  

\section{Methodology} \label{methodology}
To simplify the problem, we will focus on the electronic-only system to test the conservation of the QBD, if observables are a function of a finite number of the lowest MMST normal modes, if the MMST normal modes are constrained and finally, if truncation in MMST normal modes results in good quality dynamics. It seems unlikely that these conditions will be satisfied for a full nuclear-electronic coupled system if they are not for an electronic-only system. Therefore, the aim of this paper is to investigate the use of normal modes of the MMST $\bo{q}$ and $\bo{p}$ variables for an electronic-only system. 

The $F$-level Hamiltonian in the diabatic representation for an electronic-only system is, 
\begin{align}
    \label{hamiltonian}
    \hat{H} = \sum_{n,m=1}^F \ket{n} \bo{V}_{nm}\bra{m} \text{,}
\end{align}
where $\bo{V}$ is an $F \times F$ diabatic electronic potential matrix in the basis of electronic states, $n$, such that the MMST representation is, 
\begin{align}
    H = \frac{1}{2}\left( \bo{p}^\textrm{T}\bo{V}\bo{p} + \bo{q}^\textrm{T}\bo{V}\bo{q} - \textrm{Tr}[\bo{V}]\right) \text{,}
\end{align}
which is the third term of Eqn.~\eqref{full-MMST}, but the potential matrix is independent of nuclear position as we only have an electronic system. It is important to note that the electronic position and momenta are not themselves physical observables, as they are the mapping variables for the quantum subsystem.
The electronic-only Liouvillian is the electronic part of the decoupled Liouvillian in Eqn.~\eqref{decoupled-liouvillian},
\begin{align} \label{elecliou}
    \mathscr{L}_{\textrm{elec}} = \frac{1}{\hbar} \left[  \sum_{l=1}^{N}{\bf{p}}^{\textrm{T}}_{l} {\bf{V}} \overrightarrow{\nabla}_{{\bf{q}}_{l}} -  {\bf{q}}^{\textrm{T}}_{l} {\bf{V}} \overrightarrow{\nabla}_{{\bf{p}}_{l}} \right] \text{,}
\end{align}
which describes the motion of the electronic degrees of freedom of the system. 

The correlation functions we will evaluate are similar in form to Eqns.~\eqref{GKT_wigner-nuc} and \eqref{wigner-rt-it-nuc}, except that we need to sum over the $F$ electronic states such that,  
\begin{align} \label{GKT_wigner-main}
    C^{[N]}_{AB}(t) &= \frac{1}{Z (2\pi\hbar)^{FN}}\iint d{\bf{q}} d{\bf{p}} [e^{-\beta_N \hat{H}}\hat{A}]_{\overline{W}} \times [\hat{B}(t)]_W \text{,}
\end{align}
where, 
\begin{subequations} \label{wigner-rt-it}
\begin{align}
    &[\hat{B}(t)]_W = \prod^{N}_{j}  \int d{\boldsymbol{\mu}}'_{j} e^{i{\boldsymbol{\mu}}'_j {\bf{p}}_j/\hbar}  \braketthree{{\bf{q}}_j + {\boldsymbol{\mu}}'_j/2} { \hat{B}(t)}{{\bf{q}}_j - {\boldsymbol{\mu}}'_j/2} \text{,} \\ 
    &[e^{-\beta_N \hat{H}}\hat{A}]_{\overline{W}} = \prod^{N}_{j} \int d{\boldsymbol{\mu}}_{j} \sum_{n_{j}, m_{j}}  e^{i{\boldsymbol{\mu}}_j {\bf{p}}_j/\hbar} \braket{{\bf{q}}_{j-1} - {\boldsymbol{\mu}}_{j-1}/2}{ n_{j-1}} \nonumber \\ & \qquad \qquad \qquad \times  \braketthree{{n_{j-1}}}  {\mathcal{A} }{m_j} 
 \braket{m_j}{{\bf{q}}_j + {\boldsymbol{\mu}}_j/2} \text{,}
\end{align}
\end{subequations}
where $\mathcal{A} = (\hat{A}e^{-\beta_N\hat{H}} + e^{-\beta_N\hat{H}}\hat{A})/2$ is the symmetric operator, $\hat{B}(t) = e^{i\hat{H}t/\hbar} \hat{B} e^{-i\hat{H}t/\hbar}$ is the time-evolved operator and, $\int d\bo{q} = \prod_j^N \int d \bo{q}_j$ and likewise for $\bo{p}$.\cite{Hele2015, Hele2016} Again, the operators are linear in the sense that $\hat{A} = (\sum_j^N \hat{A}_j)/N$ and likewise for $\hat{B}$, such that they only act once in the path integral loop. However, as the operators are not a function of only position, $C_{AB}^{[N]}(t)$ may not necessarily be equal to $C_{AB}^{[N]}(-t)^*$ for a general (complex) nonadiabatic Hamiltonian.

\subsection{Electronic Normal Modes}

The electronic normal modes are defined as,
\begin{align}
    \label{normalmodetransform} 
    {\check{q}}_{kn} =  \sum_j ^N T_{jk}q_{jn} \quad \text{,} \quad \check{p}_{kn} = \sum_j ^N  T_{jk}p_{jn} \text{,}
    \end{align}
where \textit{k} is the normal mode index and \textit{n} is the electronic state index, and we sum over \textit{j} beads. The zeroth mode is related to the centroid as in Eqn.~\eqref{matsubara-modes}. We use the same transformation matrix as for normal modes of a free ring-polymer, Eqn.~\eqref{transformation}, which is similar to the transformation obtaining Matsubara modes of the nuclear variables in Matsubara dynamics.\cite{Hele2015, ceriottiEfficientStochasticThermostatting2010} We wish to make it clear that here we calculate electronic normal modes and not Matsubara modes which are defined with an additional factor of $1/\sqrt{N}$.\cite{Hele2015} We use this similar transformation as we wish to have a consistent transformation for both electronic and nuclear variables. We note that as the electronic-only system is described by a set of uncoupled harmonic oscillators at the same frequency, transforming into normal modes does not change the dynamics. 

By evaluating Hamilton's equations of motion we can obtain the exact electronic propagation equations.\cite{Cook2023, Church2018, Richardson2017} For a single bead,
\begin{align}
    \label{electprop}
    (\bo{q} + i\bo{p}) (t) = {e^{-i{\bo{V}}t}}(\bo{q}+i\bo{p})(0) \text{,}
\end{align}
where we use the shorthand $(\bo{q} + i\bo{p}) (t) \equiv \bo{q}(t) + i\bo{p}(t)$. 
For $N$ beads this becomes, 
\begin{align}
    \label{electpropmultibeads}
    (\bo{q} + i\bo{p})_j (t) = {e^{-i{\bo{V}} t}}(\bo{q}+i\bo{p})_j(0) \text{,}
\end{align}
where $j$ is the bead index and as we have an electronic-only system, the potential matrix is independent of the bead index. Hence, the electronic evolution is,
\begin{subequations}
\label{electpropderivative}
\begin{align}
    \frac{\partial}{\partial t}(\bo{q} + i\bo{p})_j (t) 
    &= -i{\bf{V}}{e^{-i{\bf{V}}t}}(\bo{q} + i\bo{p})_j(0) \\
    &= -i{\bf{V}}(\bo{q} + i\bo{p})_j (t) \text{,}
\end{align}
\end{subequations}
such that, the evolution of electronic normal modes is obtained using the transformation in Eqn.~\eqref{normalmodetransform}, as 
\begin{subequations}
\begin{align}
    \label{electpropderivative-normalmode}
    (\dot{\check{q}} + i\dot{\check{p}})_{kn} (t) &= \sum_j ^N T_{jk}(\dot{q} + i\dot{p})_{jn} (t)  \\ &=\sum_j ^N \sum_m ^F -i{\bf{V}}_{nm} T_{jk}
    (q+ip)_{jm}(t) \text{,}
\end{align}
\end{subequations}
and the back-transformation, 
\begin{align}
    \label{back-normalmodetransform} 
    {q}_{jm} = \sum_r^N T_{jr}\check{q}_{rm} \quad \text{,} \quad {p}_{jm} = \sum_r^N T_{jr}\check{p}_{rm} \text{,}
    \end{align}
is used to obtain,
\begin{align}
    \label{electpropderivative-normalmode-transformed}
    (\dot{\check{q}} + i\dot{\check{p}})_{kn} (t) &= \sum_{j,r}^N \sum_m^F -i{\bf{V}}_{nm} T_{jk} T_{jr}
    (\check{q}+i\check{p})_{rm}(t) \text{.}
\end{align}
As the ring-polymer spring matrix is a real, positive, and symmetric matrix, then the transformation matrix, $\bf{T}$, that diagonalizes it is an orthonormal matrix which results in, 
\begin{align} \label{einsumofT}
        \sum_{l}^{N} T_{lk}T_{lr} = \delta_{kr} \text{,}
    \end{align}
such that, when there is no nuclear dependence, 
\begin{align}
    \label{electpropderivative-normalmode-transformed-nonuc}
    (\dot{\check{\bo{q}}} + i\dot{\check{\bo{p}}})_{k} (t) &= -i{\bf{V}}
    (\check{\bo{q}}+i\check{\bo{p}})_{k}(t)  \text{,}
\end{align}
which is the same as for $\bo{q}$ and $\bo{p}$ [Eqn.~\eqref{electpropmultibeads}] and is separable with respect to normal modes. In Appendix~\ref{nucl-dependance}, we briefly discuss the case where there is nuclear dependence of the potential matrix. 

We can define electronic bead probabilities as, 
\begin{equation} \label{prob_time_evolved}
    \mathscr{G}_{j} (t) =  (\bo{q}-i\bo{p})^\mathrm{T}_{j}(t)(\bo{q}+i\bo{p})_{j}(t) \text{,}
\end{equation}
which is known to be conserved, such that $\mathscr{G}_{\mathrm{tot}} = \sum_{j} \mathscr{G}_j$ is also conserved. Similarly, the electronic normal mode probabilities are, 
\begin{equation} \label{nmprob}
        \check{G}_{k} (t) = (\check{\bo{q}}-i\check{\bo{p}})^\mathrm{T}_{k} (t) (\check{\bo{q}}+i\check{\bo{p}})_{k}(t) \text{,}
\end{equation}
which we show in Appendix~\ref{Conservation-of-electronic-normal-modes} is also conserved in the case of decoupled nuclear and electronic DoF.

\subsection{The Electronic-Only Liouvillian with Normal Modes} \label{Nonadiabatic-matsubara-liouvillian-in-electronic-normal-modes}
We can confirm the propagation equations by determining the Liouvillian in electronic normal modes. In Appendix~\ref{liouvillian-nm} we convert the electronic Liouvillian in Eqn.~\eqref{elecliou} into normal modes obtaining,
\begin{align} \label{l_elec_final}
    \mathscr{L}_{\textrm{elec}}
    &= \frac{1}{\hbar} \left[ \sum_{k}^N \check{{\bf{p}}}_{k}^{\textrm{T}}{\bf{V}} \overrightarrow\nabla_{\check{{\bf{q}}}_{k}} -  \check{{\bf{q}}}_{k}^{\textrm{T}}{\bf{V}} \overrightarrow\nabla_{\check{{\bf{p}}}_{k}} \right] \text{,}
\end{align}
which results in consistent propagation equations with Eqn.~\eqref{electpropderivative-normalmode-transformed-nonuc}.

A very similar derivation can be done when there is nuclear dependence to obtain Eqn.~\eqref{electpropderivative-normalmode-tensor}, as seen in Appendix~\ref{nucl-dependance}. However, we note that this is not the full coupled Liouvillian but only the electronic term. We have shown that the evolution in normal modes is very similar to the evolution in beads. In the case where there is no nuclear dependence, the evolution equations are identical and evolution in normal modes is separable. However, when there is nuclear dependence, we see the slightly more complex evolution equation in Eqn.\eqref{electpropderivative-normalmode-tensor} where the potential matrix is replaced by a tensor. 

\subsection{$N$-bead form of Correlation Function}
\subsubsection{In mapping variables}
The CF in Eqn.~\eqref{GKT_wigner-main} can be obtained in terms of mapping variables, where the real and imaginary terms are outlined in the Section \RNum{1}A of the Supplementary Material and follows a similar derivation as in references [\!~\citenum{Hele2016}] and [\!~\citenum{Chowdhury2021}], but neglecting the nuclear terms. This results in the real-time term as,
\begin{subequations}
\begin{align}
    \label{RT-overall-main}
    [\hat{B}(t)]_W 
    &= \frac{1}{2\hbar N} \sum_j^N \textrm{Tr}\left[( {\bf{C}}_{j}(t) - \hbar \mathbb{I}) {\bf{B}}_j\right]  \text{,}
\end{align}
\end{subequations}
where ${\bf{C}}_{j}(t)  = ({\bf{q}} +i{\bf{p}})_j(t)\bigotimes ({\bf{q}} -i{\bf{p}})_j^\textrm{T}(t)$ and ${\bf{B}}_{j}$ is the operator on the $j$-th bead, and the imaginary time being, 
\begin{align}
    \label{IT-overall-main}
    [e^{-\beta_N \hat{H}}\hat{A}]_{\overline{W}}
    &= \frac{2^{N(F+1)}}{\hbar^N} e^{- (\sum_j|{\bf{q}}_{j}|^2 +|{\bf{p}}_{j}|^2 )/\hbar} \textrm{Tr}\left[ \prod_j^N \mathscr{A}_j \left( {\bf{C}}_j -\frac{\hbar}{2}\mathbb{I}\right)\right] \text{,}
\end{align}
where $\mathscr{A}_j = 1/2 (\hat{A}_je^{-\beta_N\hat{H}} + e^{-\beta_N\hat{H}}\hat{A}_j) $ which is the symmetrized operator. The above is the short-hand notation to describe the following as the operators only act once in the path integral loop, 
    \begin{align}
        \label{imag-time-N}
         \textrm{Tr}\left[ \prod_j^N \mathscr{A}_j \left({\bf{C}}_j -\frac{\hbar}{2}\mathbb{I}\right)\right] \equiv &\frac{1}{N} \textrm{Tr} \left[  \sum_j \left\{ \prod_{k < j} e^{-\beta_{N}\hat{H}} \left({\bf{C}}_k -\frac{\hbar}{2}\mathbb{I}\right) \right. \right.  \nonumber \\ &\left. \left. \times  \mathscr{A}_j \left({\bf{C}}_j -\frac{\hbar}{2}\mathbb{I}\right) \right. \right.  \nonumber \\ &\left. \left. \times \prod_{n > j} e^{-\beta_{N}\hat{H}} \left({\bf{C}}_n -\frac{\hbar}{2}\mathbb{I}\right)\right\} \right] \text{,}
    \end{align}
which requires a complex series of matrix multiplications. To reduce the computational cost, we use Bell's algorithm as outlined in Appendix B of Ref.~[\!~\citenum{Hele2011}].
Hence, the overall CF is, 
\begin{align}
    \label{CF-final-main}
    C^{[N]}_{AB}(t)
    = \frac{2^{N-1}}{Z\hbar^{N+1}(\pi\hbar)^{FN} N} \iint & d{\bf{q}} d{\bo{p}}  e^{-\mathcal{G}_{\textrm{tot}}/\hbar} \textrm{Tr}\left[ \prod_j^N \mathscr{A}_j \left( {\bf{C}}_j -\frac{\hbar}{2}\mathbb{I}\right)\right] \nonumber \\ & \times  \sum_j^N \textrm{Tr}\left[( {\bf{C}}_{j}(t) - \hbar \mathbb{I}){\bf{B}}_{j}\right] \text{,}
\end{align}
where $\mathcal{G}_{\textrm{tot}} = \sum_j|{\bf{q}}_{j}|^2 +|{\bf{p}}_{j}|^2 $. The CF can be calculated by averaging over many trajectories using the Metropolis Monte-Carlo sampling outlined in Section \RNum{1}B of the Supplementary Material. If any imaginary terms arise during computation, these can be neglected as the correlation function is proven to be real in Section \RNum{1}C of the Supplementary Material.

\subsubsection{In electronic normal modes}
The CF can be written as a function of normal modes by direct transformation of $\bo{C}$, 
\begin{subequations}
\begin{align}
    \label{C-transform}
    {\bf{C}}_{j}  &= ({\bf{q}}_j +i{\bf{p}}_j)\bigotimes ({\bf{q}}_j -i{\bf{p}}_j)^\textrm{T} \\
    &= \sum_{n,m} \sum_k T_{jk} (\check{{\bf{q}}} +i\check{{\bf{p}}})_{kn} \sum_s T_{js}(\check{{\bf{q}}} -i\check{{\bf{p}}})_{sm}^\textrm{T}  \text{,}
\end{align}
\end{subequations}
such that the time-dependent part of the CF becomes, 
\begin{align}
    [\hat{B}(t)]_W = \frac{1}{2\hbar N}
     \sum_j \sum_{n,m} \Biggl\{ \sum_k^N  T_{jk} (\check{{\bf{q}}} +i\check{{\bf{p}}})_{kn} (t)  & \sum_s^N T_{js}(\check{{\bf{q}}}  -i\check{{\bf{p}}})_{sm}^\textrm{T}(t)  \nonumber \\ &  - \hbar\mathbb{I} \Biggr\}{\bf{B}}_{nm} \text{,}
\end{align}
where we consider the case where the operator is the same for all beads. We can now remove the transformation matrix with the sum over $j$ beads using Eqn.~\eqref{einsumofT},
\begin{subequations} \label{real-time-nm}
\begin{align}
    [\hat{B}(t)]_W &= \frac{1}{2\hbar N} 
     \sum_{k}^N \sum_{n,m} \left\{ (\check{{\bf{q}}} +i\check{{\bf{p}}})_{kn}(t) (\check{{\bf{q}}} -i\check{{\bf{p}}})_{km}^\textrm{T}(t) - \hbar\mathbb{I} \right\}{\bf{B}}_{nm} \\
    &= \frac{1}{2\hbar N}  \sum_k^N \textrm{Tr}\left[( \check{{\bf{C}}_{k}}(t) - \hbar \mathbb{I}){\bf{B}}_{j}\right] \text{,}
\end{align}
\end{subequations}
where $\check{{\bf{C}}_{k}} = (\check{{\bf{q}}} +i\check{{\bf{p}}})_k \bigotimes(\check{{\bf{q}}} -i\check{{\bf{p}}})^\textrm{T}_k $. One can sample $\bo{q}$ and $\bo{p}$, calculate the zero-time term, Eqn.~\eqref{imag-time-N}, and then transform to and propagate the normal modes to calculate the time-dependent term, Eqn.~\eqref{real-time-nm}. 

If $\bo{B}(t)$ is the population of state 1 for a two-level system with $N$ beads,
\begin{align}
\label{elec-prob-state-1-all-modes-needed}
     [\hat{B}(t)]_W
     &= \frac{1}{2\hbar N} \sum_{k}^N \left\{ \check{q}^2_1 (t)+ \check{p}^2_1(t) - \hbar \right\} \text{,}
\end{align}
which is dependent on all $k$ normal modes. From this we can see that the observable is not a function of only a finite number of the lowest normal modes as all are required in the sum. However, this does not necessarily mean that the higher normal modes will contribute significantly to the dynamics.

\subsection{Conservation of the quantum Boltzmann distribution}
    
In Appendix~\ref{cons-qbd}, we prove QBD conservation for both a single-bead and multi-bead calculation for a single trajectory. In the multi-bead case, acting the Liouvillian brings down a $\pm\bo{V}$ either side that bead's term, as seen in Figure~\ref{fig:liouvillian-acting}. This means that the terms only cancel out when completing the path integral loop, so all beads are required for the QBD to be conserved. The multi-bead proof can be used to prove that a calculation with all the normal modes included will also conserve the distribution. A back-transformation can be performed to obtain the bead form from the normal mode calculation, and this is equivalent to including all the beads. This also suggests that truncating in beads or normal modes will not conserve the QBD for a single trajectory, likely a consequence of the observable not being a function of only a finite number of the lowest normal modes and that the transformation matrix into normal modes cannot be removed when truncating. We investigate this computationally later in this article and leave further algebraic investigation of this as future work.

\section{Results}\label{results}
\begin{figure}[H]
    \centering
    \includegraphics[width=\linewidth]{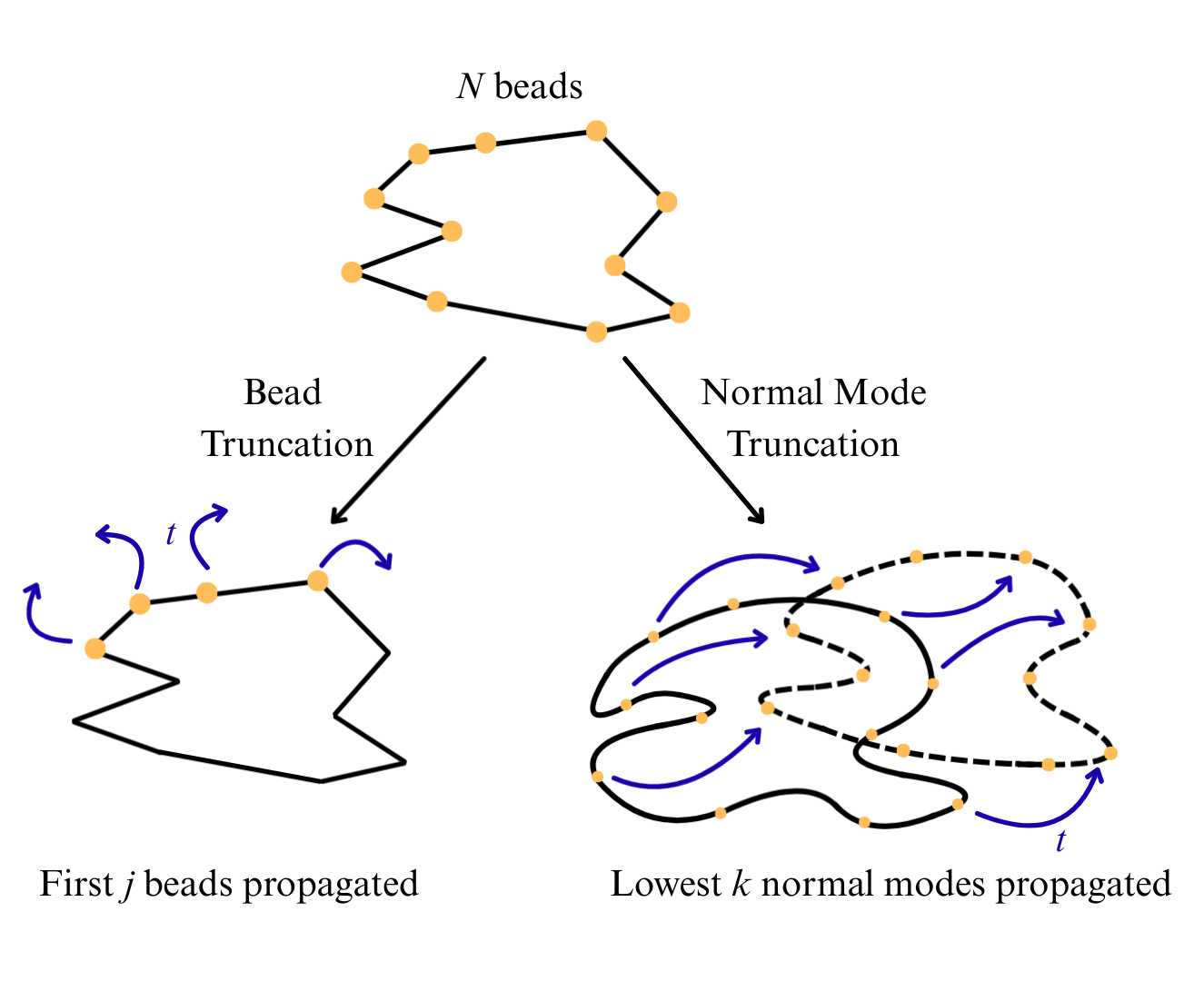}
    \caption{\label{fig:truncating} Schematic diagram illustrating bead and normal mode truncation of the path integral, where the blue arrows indicate time propagation. The dashed path integral is time-evolved. Bead truncation is where the first $j$ beads are propagated. Normal mode truncation is where the lowest $k$ normal modes are propagated. All unpropagated beads and normal modes are left at their initial time ($t=0$) values.}
\end{figure}

For simplicity, we consider only the electronic system in our calculation, as mentioned earlier. We think it is very unlikely that a general QBD-conserving method will be found which does not conserve the QBD for the electronic-only system. 
Here, we discuss the first electronic state population auto-CF and the conservation of the QBD when truncating in beads and normal modes. We also provide additional plots that aid our understanding of how each bead and normal mode contributes to the calculations. When truncating in the bead representation ($\bo{q}$ and $\bo{p}$), we propagate the first $j$ beads and keep the remaining $j+1 \to N$ beads at their initial time values. Likewise, when truncating in normal modes ($\check{\bo{q}}$ and $\check{\bo{p}}$), we propagate the lowest $k$ normal modes and all higher normal modes are kept at their initial time values. We can do this as the normal modes are separable for the electronic-only system, seen in Eqn.~\eqref{electpropderivative-normalmode-transformed-nonuc}. A schematic for this can be seen in Figure \ref{fig:truncating}.

We model the potential energy matrix for a two-level system as,
\begin{align}
    \bo{V} = \begin{bmatrix}
        \alpha & \Delta \\
        \Delta & -\alpha
    \end{bmatrix} \text{,}
\end{align}
where the asymmetry, $\alpha = 1$ and the electronic coupling, $\Delta = 1$, such that we model an asymmetric system. In all cases, we utilised reduced units such that $\beta = 1$. All the results shown here are for a 8 bead calculation where we have also transformed into the 8 normal modes and truncated to the lowest 1, 3 and 5 beads and normal modes. Results for a symmetric system where $\alpha = 0$ with $N = 8$ and $N = 4$ can be seen in the Supplementary Material (Figures S.1--6) and for an asymmetric potential where $N = 4$ (Figures S.7--9). We note that an 8 bead calculation considering 5 normal modes is significantly outside the Matsubara limit (where the number of beads is far greater than the number of normal modes), however, larger numbers of beads makes convergence challenging due to the Metropolis Monte-Carlo sampling and we find $N=8$ provides us with enough information to probe adequately the properties of the MMST normal modes. 

\subsection{Autocorrelation function}
\begin{figure}[H]
    \centering
    \includegraphics[width=\linewidth]{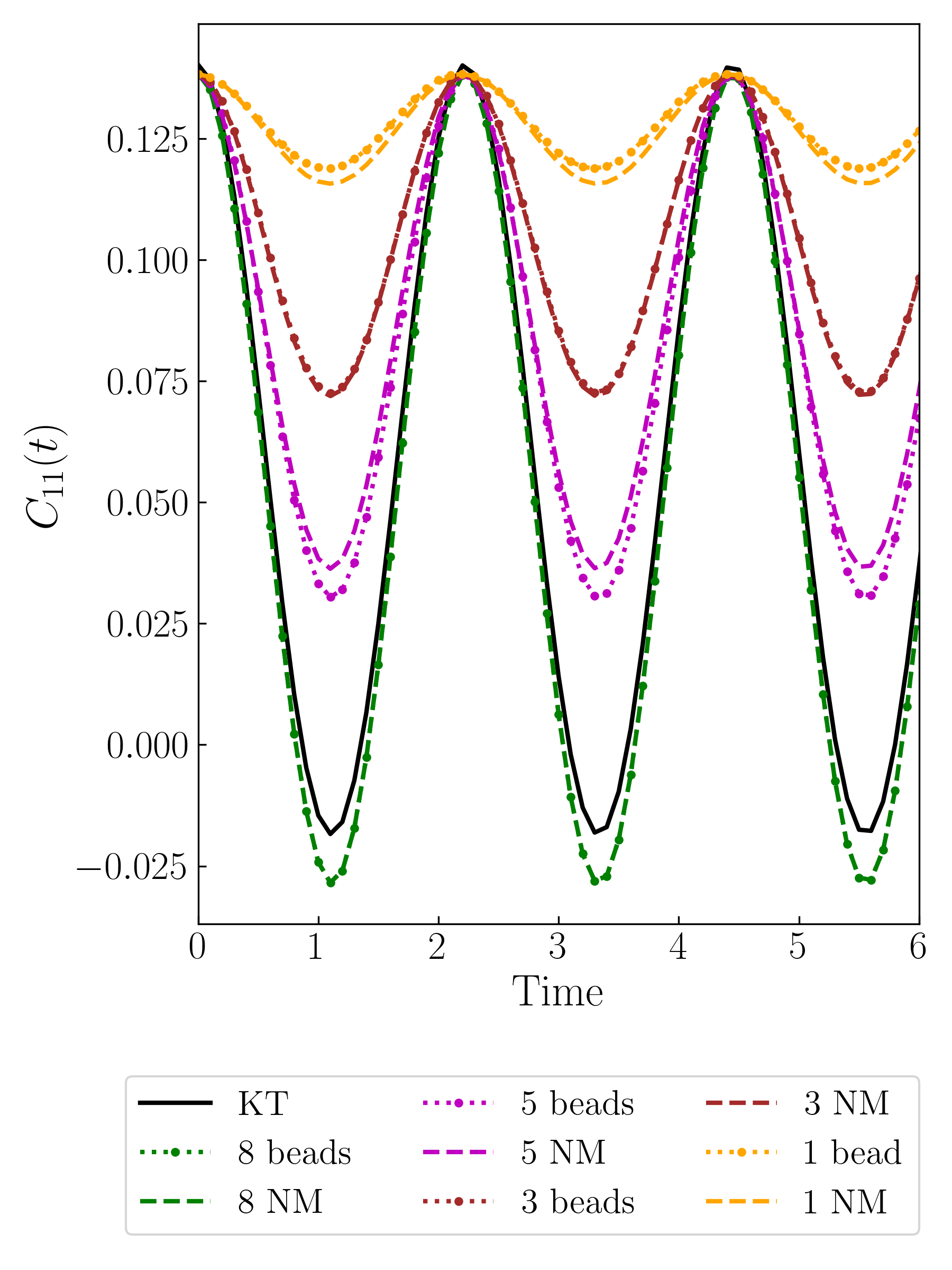}
    \caption{\label{fig:steps}Correlation function for the electronic population of the first state with the exact Kubo-Transformed result (KT, black) compared with a full 8 bead calculation with truncation in both beads (dotted circle) and normal modes (NM, dashed) for; 8 (green), 5 (magenta), 3 (red) and 1 (orange). When truncating, for example with 5 beads/normal modes, the first 5 beads or the lowest 5 normal modes are propagated, with the remaining beads/normal modes kept at their initial time values. While the result improves with more beads/normal modes included, all need to be included to obtain an accurate result.}
\end{figure}

Firstly, we calculate the $C_{11}(t)$ CF using a full 8 bead and normal mode calculation, and when truncating in beads or normal modes, shown in Figure \ref{fig:steps}. When truncating, we keep the truncated beads or normal modes at the zero-time sampled value. We see that the more beads/normal modes that are included, the result systematically becomes more accurate compared to the exact KT result (solid black line), as seen in Figure \ref{fig:steps}. This plot demonstrates numerically what we have shown algebraically in Eqn.~\eqref{elec-prob-state-1-all-modes-needed} that the observable is a function of not only the lowest normal modes but all normal modes, and that the higher normal modes do still contribute significantly to the dynamics. Each set of bead and normal mode results converge on similar values, highlighted by the $N=4$ results in the Supplementary Material (Figure S.2) where convergence is easier to obtain, emphasizing that the transformation does not change the dynamics for the electronic-only system. 

\subsection{Quantum Boltzmann Distribution Conservation}
We can test conservation of the QBD for a single trajectory by plotting the Boltzmann term, $\textrm{Tr}\left[ \prod_j^N e^{-\beta_N \bo{V}}\left( {\bf{C}}_j(t) -\frac{\hbar}{2}\mathbb{I}\right)\right]$, where we evolve $\bo{q}$ and $\bo{p}$ and calculate $\bo{C}_{j}(t)$. To test the conservation in normal modes, we back-transform into beads from the propagated $\check{\bo{q}}$ and $\check{\bo{p}}$ and calculate the Boltzmann term using the same formula. We do this as the trace of a product of matrices means that the transformation matrix cannot be easily eliminated, and it is computationally easier to calculate this term in beads. When we truncate in beads/normal modes, we keep the higher beads/normal modes at their initial time values before doing the calculation as mentioned earlier.

\begin{figure}[H]
        \centering
        \includegraphics[width=\linewidth]{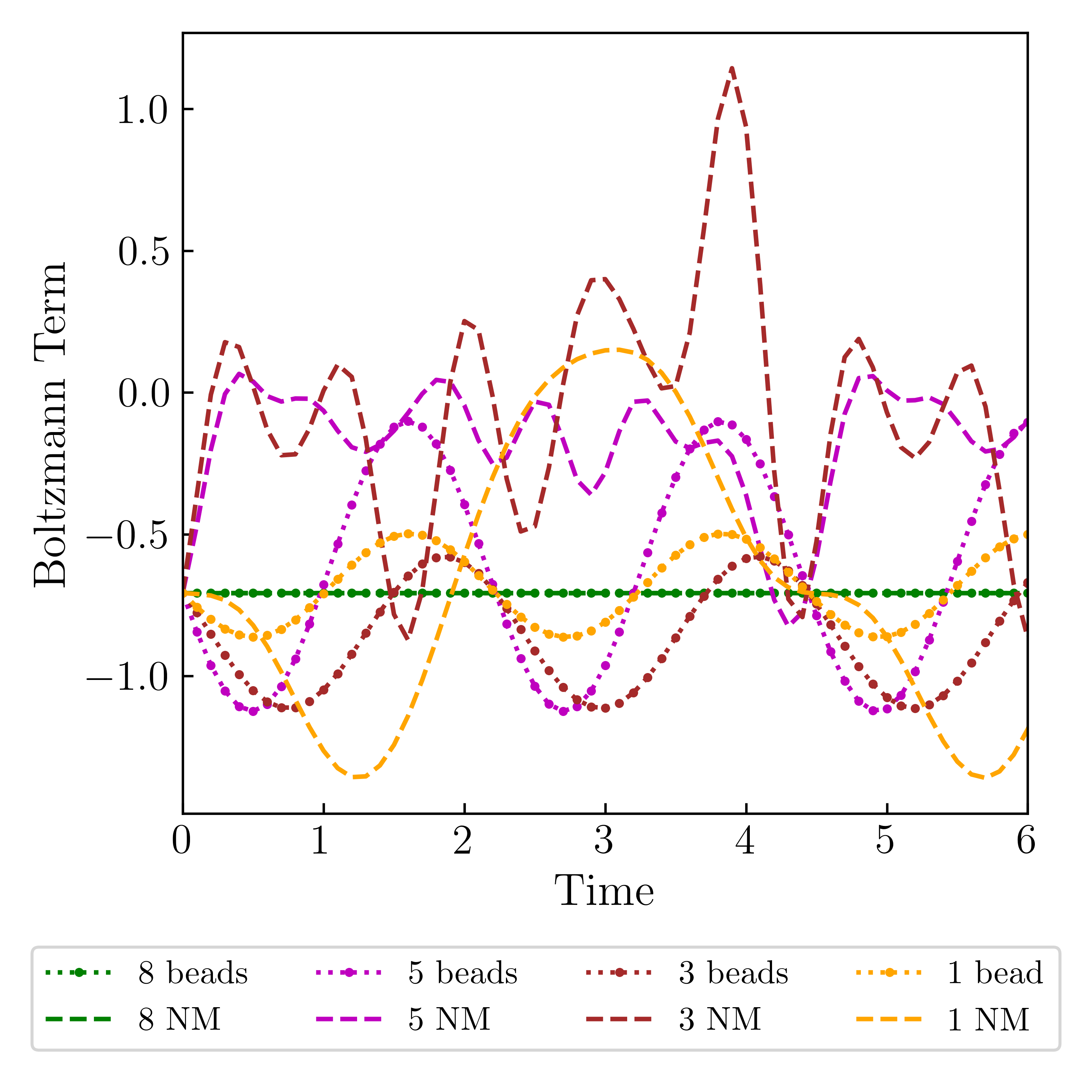}
        \caption{\label{fig:singletrajconsen}The propagated Boltzmann term against time for a single trajectory with a full 8 bead calculation with truncation in both beads (dotted circle) and normal modes (dashed) for; 8 (green), 5 (magenta), 3 (red) and 1 (orange). The 8 beads and normal modes are the only lines that are flat, indicating QBD conservation.}
\end{figure}

The single trajectory results in Figure~\ref{fig:singletrajconsen} agree with the algebra in Appendix~\ref{cons-qbd}. The full 8 bead and normal mode calculations conserve the QBD, as seen by the flat line (green dotted and dashed), for a single trajectory. The dotted and dashed green lines are the same such that it is only possible to see the dashed green line in this plot. Any truncation in both the beads and normal modes results in the line oscillating, indicating that this term is not conserved. The energy of a single trajectory, $\textrm{Tr}\left[ \prod_j^N \bo{V}\left( {\bf{C}}_j(t) -\frac{\hbar}{2}\mathbb{I}\right)\right]$, can also be tested for conservation (Figure S.3 in the Supplementary Material) where we see the same trend as for the Boltzmann term. 

We can test the conservation of an ensemble of trajectories by calculating the expectation value through the evolved CF, $C_{\mathbb{I}1}$. Although algebraically we can show that all beads are required to obtain QBD conservation for a single trajectory, this is challenging to determine in normal modes due to the inability to transform between $\bo{C}_j$ in beads and $\check{\bo{C}}_k$ in normal modes when truncating. This makes it unlikely that the QBD will be conserved when truncating in normal modes. While we can show that including all normal modes conserves the QBD, we leave a full description of the effect of truncation algebraically as further work. The numerical results for a single trajectory show that truncating in normal modes does not conserve the QBD. However, this does not mean there will not be an averaging effect that results in conservation for many trajectories.

\begin{figure}[H]
        \centering
        \includegraphics[width=\linewidth]{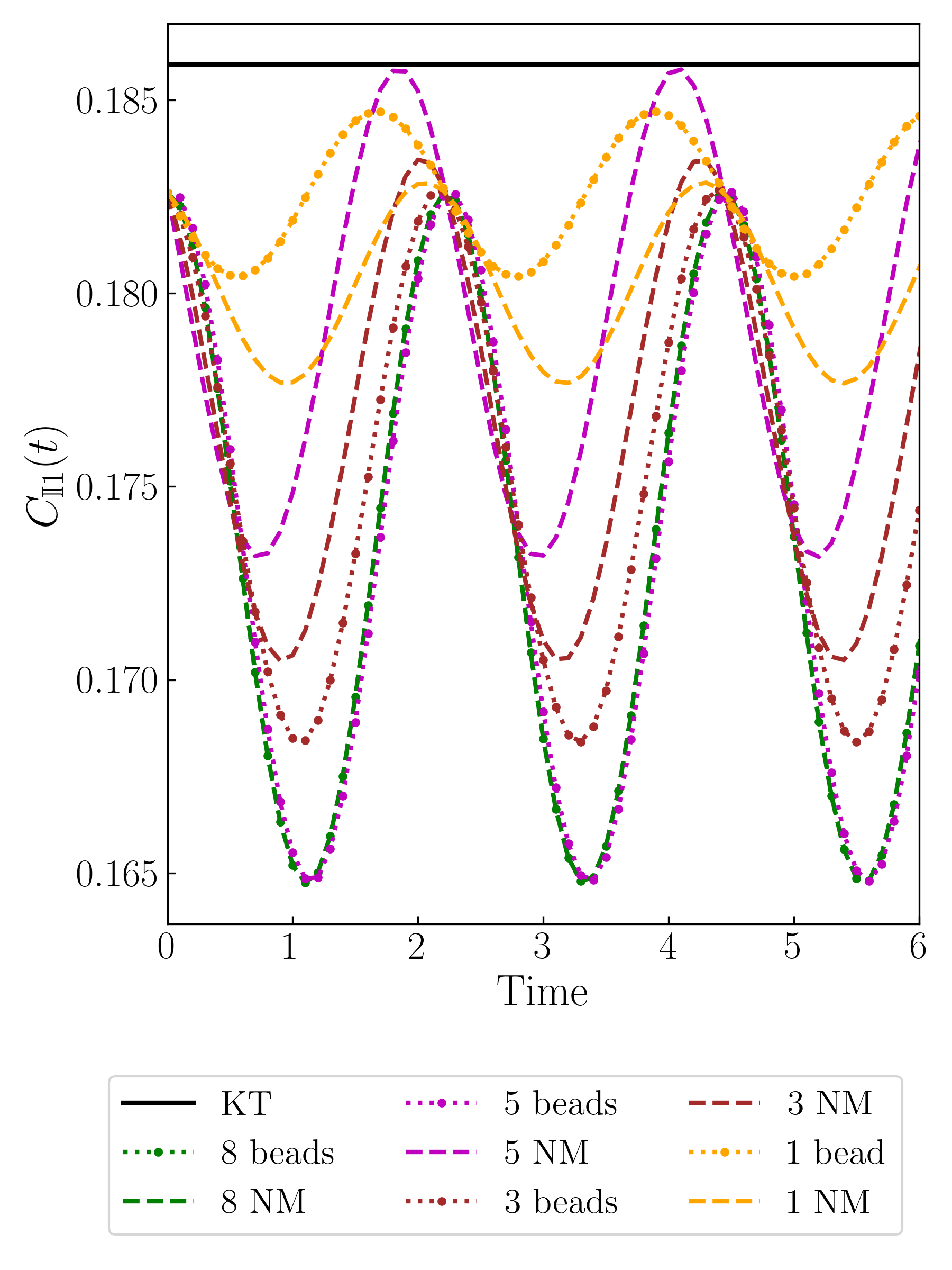}
        \caption{\label{fig:steps conserve}Correlation function for the conservation of electronic population of the first state with the exact Kubo-Transformed result (KT, black) compared with a full 8 bead calculation with truncation in both beads (dotted circle) and normal modes (NM, dashed) for; 8 (green), 5 (magenta), 3 (red) and 1 (orange). All lines oscillate quite close to the Kubo-Transformed result (note the y-axis scale) so there appears to be an averaging effect that conserves the QBD for an ensemble of trajectories.}
\end{figure}

In Figure~\ref{fig:steps conserve}, the CF is plotted for 8 beads/normal modes and truncated calculations. The convergence of the sampling to obtain a straight line requires many trajectories and full convergence is difficult to obtain. We see that all lines appear to have similar oscillations with not much difference between them. It appears that averaging over many trajectories results in the appearance of conservation even for the truncated bead/normal mode calculations (noting the scale of the y-axis in Figure~\ref{fig:steps conserve}), even though numerically they do not conserve the distribution for a single trajectory. 

\begin{figure}[H]
        \centering
        \includegraphics[width=\linewidth]{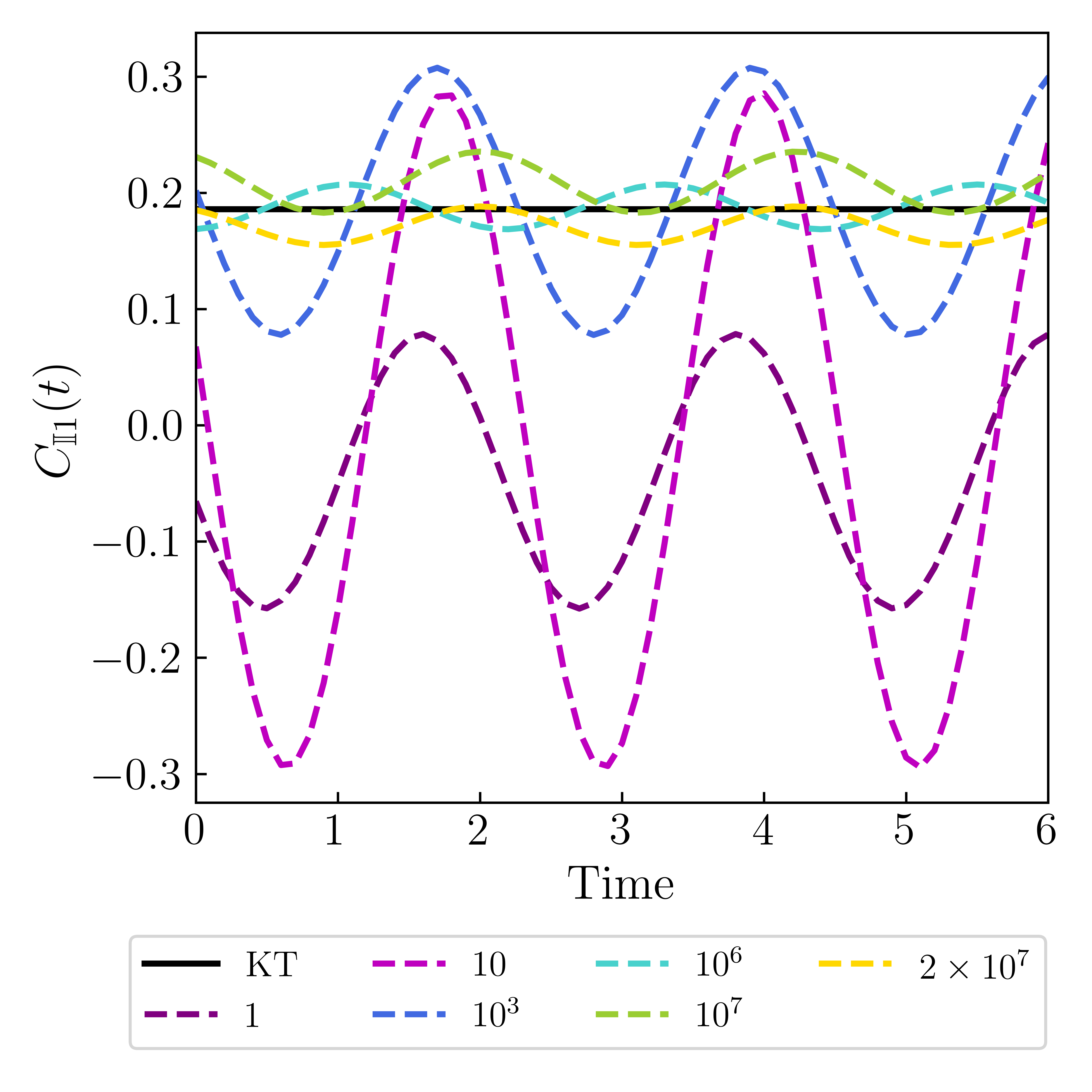}
        \caption{\label{fig:moretrajcons}Correlation function for the conservation of electronic population of the first state with the exact Kubo-Transformed result (KT, black) compared with a full 8 bead calculation with different numbers of trajectories; 1 (purple), 10 (magenta), $10^3$ (blue), $10^6$ (light blue), $10^7$ (green) and $2\times 10^7$ (yellow). A general trend is observed where increasing the number of trajectories improves the accuracy of the correlation function and the amplitude of oscillations decreases.}
\end{figure}

In Figure~\ref{fig:moretrajcons}, we show how including more trajectories converges the calculation towards a flat line for 8 beads/normal modes. The plot shows how increasing the number of trajectories initially increases the size of oscillations but when a suitable amount of trajectories are included such that a large enough amount of phase space is sampled, in this case 1 million trajectories, the line converges upon the expected value. We see the same trend for truncated calculations in beads/normal modes, which can be seen in Figure S.10 the Supplementary Material.

\subsection{MMST normal mode distribution}
\begin{figure*}[ht]
        \centering
        \includegraphics[width=0.9\linewidth]{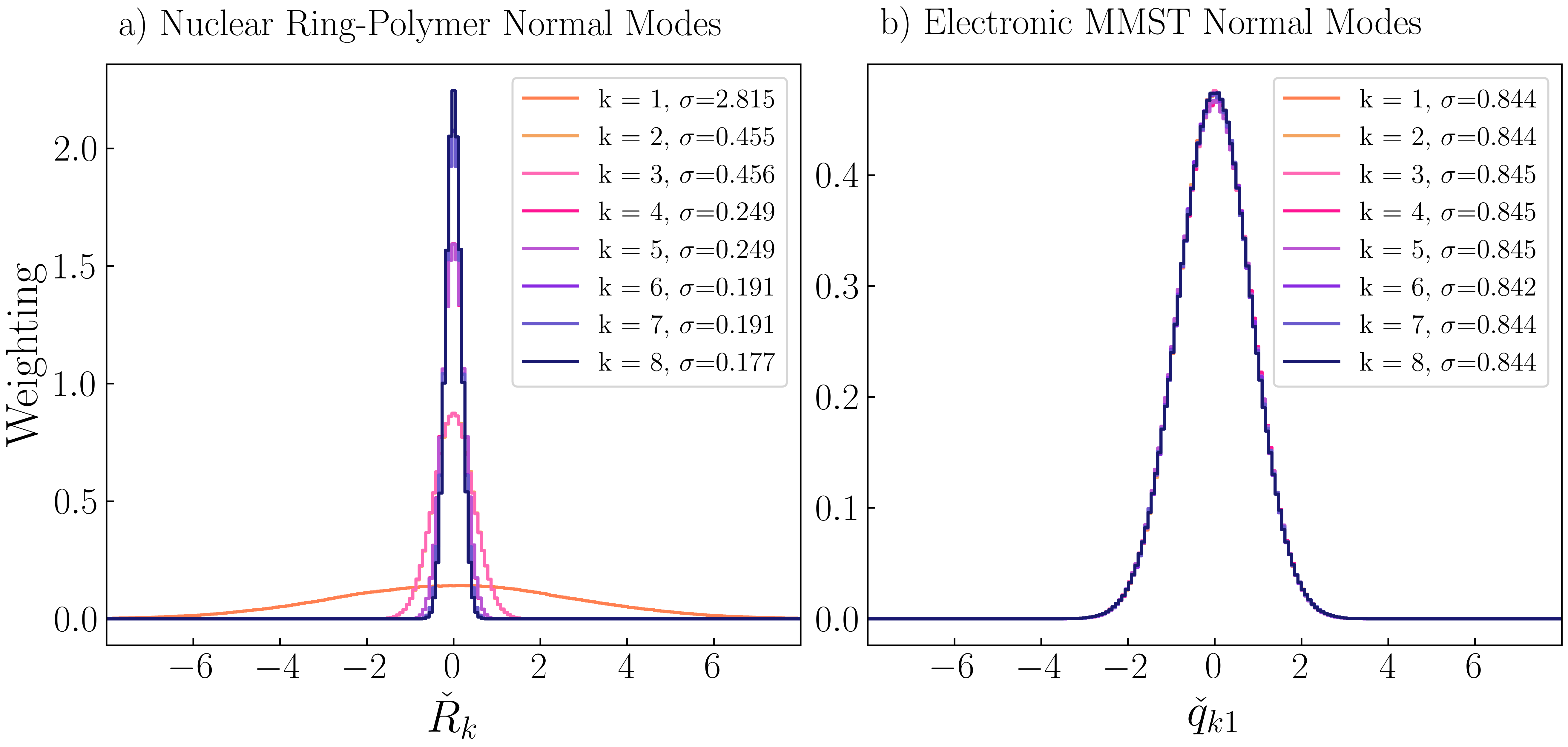}
        \caption{\label{fig:normalmodes} Histogram of a) ring-polymer nuclear normal modes and b) MMST normal modes of $\bo{q}$ for the first electronic state (right plot) with the standard deviation in the legend. We see that the constraint on the ring-polymer normal modes narrows and elongates the distribution. The lack of narrowing in the MMST normal mode distributions indicates that these are not constrained.}
\end{figure*}

To try and understand why all the normal modes need to be included to obtain an accurate CF, we look at the distribution of MMST normal modes. In Figure~\ref{fig:normalmodes}, we compare the distribution of MMST normal modes of $\bo{q}$ (for the first electronic state using the asymmetric potential outlined earlier) with nuclear position normal modes for a ring-polymer in a harmonic potential, $V =  R^2/2$. For the ring-polymer normal modes (Figure~\ref{fig:normalmodes}a), we see the distribution significantly narrows and becomes taller as the mode index increases. This is as the normal modes are constrained by the springs between the beads. In Matsubara dynamics, the higher, truncated (unevolved) normal modes can be integrated out by performing a contour integral which is valid at $t=0$.\cite{heleCommunicationRelationCentroid2015, Hele2015} This results in spring-like terms between the higher normal modes and so the truncated normal modes will be constrained in a similar manner to the ring-polymer normal modes seen here.\cite{heleCommunicationRelationCentroid2015, Hele2015} However, for the MMST normal modes (Figure~\ref{fig:normalmodes}b), the distributions are all the same, due to a lack of springs between the electronic DoF, and we do not see any narrowing.

In the Supplementary Material, we investigate several other electronic population metrics involving MMST variables as seen in Figures S.11-13. However, none of these metrics resulted in a narrowing of the normal mode distributions. This shows that normal modes of MMST variables, and of the associated population metrics based on MMST variables, are not optimal metrics for constructing a QBD-conserving method with a Matsubara approach. 

\subsection{Summary}
\begin{table} [H]
\caption{\label{table1}Summary of the results in relation to our earlier criteria for a desirable electronic normal modes metric. The right-hand column contains the main results of this article, which as far as we are have not been published before.}
\begin{center}
    \begin{tabular}{l|c|c}
      \toprule 
      \textbf{Criterion} & \makecell[c] {\textbf{Nuclear} \\ \textbf{Matsubara Modes }} & \makecell[c] {\textbf{MMST} \\ \textbf{Electronic Modes }} \\
      \midrule 
      \rowcolor[HTML]{C0C0C0} \makecell[l]{\parbox{3.3cm}{ Observable function \\ of finite number of NM}} & \cmark & \xmark \\
      Constrained higher NM & \cmark & \xmark  \\
      \rowcolor[HTML]{C0C0C0} \makecell[l]{\parbox{3.3cm}{Truncated single trajectory QBD conservation}} & \cmark & \xmark \\
      \makecell[l]{\parbox{3.3cm}{Truncated ensemble \\ QBD conservation}} & \cmark & \cmark (numerically)  \\
      \rowcolor[HTML]{C0C0C0} \makecell[l]{\parbox{3.3cm}{Accurate dynamics \\ upon truncation}} & \cmark & \xmark  \\
      \bottomrule 
    \end{tabular}
\end{center}
\end{table}

Overall, we find that electronic normal modes directly calculated from $\bo{q}$ and $\bo{p}$ does not simplify the problem enough to calculate the CF only using a finite number of the lowest normal modes. Any truncation in normal modes results a loss of conservation of the QBD for a single trajectory but there does appear to be an averaging effect for an ensemble of trajectories that results in conservation. We have tabulated our findings in relation to the criteria for a desirable electronic state metric (set out in Section~\ref{backgroundtheory}) in comparison to nuclear Matsubara modes in Table~\ref{table1}, which summarizes the main results of this article. 

Although it appears that the MMST representation is not optimal for describing electronic normal modes, due to the lack of springs between electronic DoF, this does not mean that no other metric exists which might be suitable. Truncating in electronic normal modes described by some other metric may still allow us to obtain both accurate CFs from a finite number of normal modes and conservation of the QBD. We leave investigation of alternative metrics as future research.

\section{Conclusions} \label{conclusions}
In this article, we have considered in detail for the first time the electronic path integral normal modes of the position and momentum variables of the Meyer--Miller--Stock--Thoss (MMST) representation of nonadiabatic dynamics.
We have discussed the use of electronic normal modes calculated through the `obvious' method of directly transforming $\bo{q}$ and $\bo{p}$ from the MMST mapping Hamiltonian, as is done for the nuclear variables in Matsubara dynamics. Whilst  we could have considered an alternative transformation matrix or a different metric to take normal modes of, we believe the study of MMST normal modes with the same transformation as for ring-polymer normal modes is of significant interest. The transformation is well understood and utilised in Matsubara dynamics,\cite{ceriottiEfficientStochasticThermostatting2010, Hele2015} and the proposed NA-Mats method is constructed in MMST variables but does not take electronic normal modes or generally conserve the QBD.\cite{Chowdhury2021}  

We do not find that the observables are a function of only a finite number of the lowest normal modes, that the distribution of the higher normal modes significantly narrows, that the QBD is conserved for a single trajectory when truncating in either beads or normal modes, or that accurate dynamics are obtained from truncation. The ensemble calculations are also particularly challenging to converge with a 2-level electronic-only system and there is an averaging effect that results in QBD conservation for all tested truncations. The lack of narrowing of the distribution arises from the lack of a Boltzmann constraint or penalty in these electronic DoF, unlike in the conventional nuclear ring-polymer where the imaginary-time path integral leads to `springs' which results in a narrowing in the distribution of higher normal modes. These results suggest that normal modes of the MMST variables may not be the best metric to derive a nonadiabatic dynamics method satisfying the four criteria outlined in the introduction. However, this does not mean that we cannot find a nonadiabatic Matsubara method that will conserve the QBD distribution, for both a single and ensemble of trajectories, but rather that it requires a different metric from the MMST $\bo{q}$ and $\bo{p}$. Whilst here we consider a few simple metrics constructed from MMST electronic variables, this is by no means a complete search and we intend to consider additional metrics as future research, including spin-mapping,\cite{Runeson2019, Runeson2020, Bossion2021, bossion_non-adiabatic_2022} which may result in a QBD conserving method. 

This also suggests that were one to construct a general nonadiabatic Matsubara method based on MMST variables that truncated in both the electronic and nuclear normal modes it would be unlikely to conserve the QBD (at least for a single trajectory) and its dynamics would be unlikely to be a good approximation to the exact quantum result. These results may also explain why, despite the variety of nonadiabatic dynamics methods using the MMST representation, we are unaware of any such method which uses the MMST Hamiltonian to propagate dynamics and also conserves the QBD.\cite{Meyer1979, Stock2005,Church2018, Richardson2017, Sun1998, Sun1998b, Ananth2007, Kelly2012, saller_improved_2020, Huo2011, kirrander_ehrenfest_2020}


\section*{Supplementary Material}
The supplementary material contains additional algebra for: the CF derivation in mapping variables, the partition function and sampling details and, proof that the CF is real. We provide additional figures for a symmetric potential with $N = 4$ and $N = 8$ beads. For the asymmetric potential we present: the $N = 4$ results, $N=8$ ensemble convergence results with different truncations and, the distributions of alternative electronic population metrics (based on MMST variables).  

\section*{Acknowledgements}
TJHH acknowledges a Royal Society University Research Fellowship URF\textbackslash R1\textbackslash 201502. LEC acknowledges a University College London studentship. We thank Stuart Althorpe for interesting conversations during this research and James Rampton for comments on the manuscript.

\section*{Data Availability}
The data that support the findings of this study are openly available in UCL Research Data Repository at https://doi.org/10.5522/04/c.7993391, reference number 7993391.


%
%

%

\appendix
    \renewcommand{\thesubsection}{\Alph{subsection}}
    \section*{Appendices}
    \addcontentsline{toc}{section}{Appendices}
    
    \subsection{Electronic Mode Conservation} \label{Conservation-of-electronic-normal-modes} 
    The electronic state probabilities are defined in Eqn.~\eqref{prob_time_evolved} and the total electronic probability is,
    \begin{align} \label{tep} 
        \mathscr{G}_{\textrm{tot}} (t) &= \sum_{j}^N \mathscr{G}_{j} (t) =  \sum_{j}^N \sum_{n}^F {{q}}_{jn}^{2}(t) + {{{p}}^{2}_{jn}}(t) \text{,}
    \end{align}
    which as the sum of conserved values will also be conserved.
    Likewise, the electronic normal mode probabilities are defined in Eqn.~\eqref{nmprob} where using the transformation in Eqn.~\eqref{normalmodetransform},
    \begin{equation} \label{ssep1}
        \check{G}_{k} (t) =  \sum_{j,l}^N \sum_{n}^F T_{lk}(q-ip)_{ln}(t) \ T_{jk}(q+ip)_{jn}(t) \text{,}
    \end{equation}
    and using the evolution, Eqn.~\eqref{electpropmultibeads}, we obtain
    \begin{equation} \label{enmprop2}
        \check{G}_{k} (t) =  \sum_{j,l}^N \sum_{n,r,m}^F T_{lk}{(q-ip)_{lr}(0)e^{+i {\bf{V}}t}}_{lnr} \ {e^{-i {\bf{V}}t}}_{jnm}T_{jk} (q+ip)_{jm}(0) \text{.}
    \end{equation}
    In the case where there is no nuclear dependence of the potential matrix,
    
    \begin{subequations}
    \begin{align}
        \label{enmpkronecker} 
        \check{G}_{k} (t) &=  \sum_{j,l}^N \sum_{r,m}^F T_{lk}(q - ip)_{lr}(0)\delta_{rm} T_{jk} (q+ip)_{jm}(0)\\ 
        &= \check{G}_{k} (0) \text{,}
    \end{align}
    \end{subequations}
    such that the electronic probability of a normal mode is conserved. 
    
    However, if the potential matrix has nuclear dependence, then Eqn.~\eqref{enmprop2} will not become Eqn.~\eqref{enmpkronecker} as, 
    \begin{align}
        \label{identity_with_nuc}
        \sum_{n}^F e^{i{\bf{V}}t}_{lnr}e^{-i{\bf{V}}t}_{jnm} = \delta_{rm} \quad \textrm{if} \quad l = j \text{,}
    \end{align}
    and in general, this is not true if $l \neq j$.
    Hence, this only holds for an electronic-only system where there is no nuclear dependence.

    \subsection{Electronic Liouvillian in Normal Modes}
    \label{liouvillian-nm}
    We aim to convert Eqn.~\eqref{elecliou}, rewritten below for convenience, 
    \begin{align}
        \mathscr{L}_{\textrm{elec}} = \sum_{l=1}^{N} \frac{1}{\hbar} \left[ {\bf{p}}^{\textrm{T}}_{l} {\bf{V}} \overrightarrow{\nabla}_{{\bf{q}}_{l}} -  {\bf{q}}^{\textrm{T}}_{l} {\bf{V}} \overrightarrow{\nabla}_{{\bf{p}}_{l}} \right]\text{,}
    \end{align}
    into normal modes. 
    We can express ${\bf{p}}^{\textrm{T}}_{l} {\bf{V}} \overrightarrow{\nabla}_{{\bf{q}}_{l}}$ as, 
    \begin{align} \label{electronicrecasting}
        {\bf{p}}^{\textrm{T}}_{l} {\bf{V}} \overrightarrow{\nabla}_{{\bf{q}}_{l}} = \sum_{n,m}^F p_{ln}{\bf{V}}_{nm} \left. \frac{\partial}{\partial q_{lm}}  \right|_{q_{j \neq l, s \neq m}} \text{,}
    \end{align}
    and utilise the back transformation, Eqn.~\eqref{back-normalmodetransform}, to obtain normal modes, 
    \begin{align}
        {\bf{p}}^{\textrm{T}}_{l} {\bf{V}} \overrightarrow{\nabla}_{{\bf{q}}_{l}} = \sum_{k}^N \sum_{n,m}^F  T_{lk}\check{p}_{kn}{\bf{V}}_{nm} \left. \frac{\partial}{\partial q_{lm}}  \right|_{q_{j \neq l, s \neq m}} \text{,}
    \end{align}
    From the definition of a total derivative we obtain, 
    \begin{align}
        \left. \frac{\partial f({\check{{\bf{q}}}})}{\partial{q_{lm}}} \right|_{q_{j \neq l, s \neq m}} = \sum_{r}^N \left. \frac{\partial f({\check{{\bf{q}}}})}{\partial \check{q}_{rm}}  \right|_{\check{q}_{s \neq m, w \neq r}} \left. \frac{\partial \check{q}_{rm}}{\partial{q_{lm}}} \right|_{q_{j \neq l, s\neq m}} \text{,}
    \end{align}
    where $ f({\check{{\bf{q}}}})$ is a ghost function, such that, using Eqn.~\eqref{normalmodetransform}, 
    \begin{align}
        \left. \frac{\partial f({\check{{\bf{q}}}})}{\partial{q_{lm}}} \right|_{q_{j \neq l, s \neq m}} = \sum_{r}^N \left. \frac{\partial f({\check{{\bf{q}}}})}{\partial \check{q}_{rm}}  \right|_{\check{q}_{s \neq m, w \neq r}} \left. \sum_{u}^N \frac{\partial T_{ru}q_{um}}{\partial{q_{lm}}} \right|_{q_{j \neq l, s\neq m}} \text{,}
    \end{align}
    where, 
    \begin{align}
         \sum_{u}^N \left. \frac{ \partial T_{ru}q_{um}}{\partial{q_{lm}}} \right|_{q_{j \neq l, s\neq m}} = 
              T_{lr} \delta_{lu} \text{.}
    \end{align}
    We can now express Eqn.~\eqref{electronicrecasting} as 
    \begin{align} \label{electronicrecastedp}
        {\bf{p}}^{\textrm{T}}_{l} {\bf{V}} \overrightarrow{\nabla}_{{\bf{q}}_{l}} = \sum_{k,r}^N \sum_{n,m}^F T_{lk}\check{p}_{kn}{\bf{V}}_{nm} T_{lr} \left. \frac{\partial}{\partial \check{q}_{rm}}  \right|_{\check{q}_{s \neq m, w \neq r}} \text{,}
    \end{align} 
    and likewise for the second term,
    \begin{align} \label{l_elec_2}
        \mathscr{L}_{\textrm{elec}} = \sum_{l=1}^{N} \frac{1}{\hbar} & \left[ \sum_{k,r}^N \sum_{n,m}^F  T_{lk}\check{p}_{kn}{\bf{V}}_{nm} T_{lr} \left. \frac{\partial}{\partial \check{q}_{rm}}  \right|_{\check{q}_{s \neq m, w \neq r}} \nonumber \right. \\ & \left. -  T_{lk}\check{q}_{kn}{\bf{V}}_{nm} T_{lr} \left. \frac{\partial}{\partial \check{p}_{rm}}  \right|_{\check{p}_{s \neq m, w \neq r}} \right] \text{.}
    \end{align}
    This can be simplified by using Eqn.~\eqref{einsumofT} to obtain, 
    \begin{subequations}
    \begin{align} \label{l_elec_3}
        \mathscr{L}_{\textrm{elec}} &= \frac{1}{\hbar} \left[ \sum_{k}^N \sum_{n,m}^F \check{p}_{kn}{\bf{V}}_{nm} \left. \frac{\partial}{\partial \check{q}_{km}}  \right|_{\check{q}_{s \neq m, w \neq k}} -  \check{q}_{kn}{\bf{V}}_{nm} \left. \frac{\partial}{\partial \check{p}_{km}}  \right|_{\check{p}_{s \neq m, w \neq k}} \right] \\
        &= \frac{1}{\hbar} \left[ \sum_{k}^N \check{{\bf{p}}}_{k}^{\textrm{T}}{\bf{V}} \overrightarrow\nabla_{\check{{\bf{q}}}_{k}} -  \check{{\bf{q}}}_{k}^{\textrm{T}}{\bf{V}} \overrightarrow\nabla_{\check{{\bf{p}}}_{k}} \right] \text{.}
    \end{align}
    \end{subequations}
    It is easy to show that acting this Liouvillian on $(\check{q}+i \check{p})_{km}$ (where $\hbar = 1$) is consistent with the propagation equation in Eqn.~\eqref{electpropderivative-normalmode-transformed-nonuc}. 

    \subsection{Nuclear Dependence}
    \label{nucl-dependance}
    In the more general case where there is nuclear dependence of the potential matrix, denoted by a subscript $l$, then the bead evolution in Eqn.~\eqref{electpropderivative} becomes, 
    \begin{align}
        (\dot{q}+ i\dot{p})_{ln} (t)  
        &= -i{\bf{V}}_{lnm}(q + ip)_{lm} (t) \text{,}
    \end{align}
    and the normal mode evolution is then,
    \begin{align}
        \label{electpropderivative-normalmode-transformed-nuc-dep}
        (\dot{\check{q}} + i\dot{\check{p}})_{kn} (t) &= -i{\bf{V}}_{lnm} T_{lk} T_{lr}
        (\check{q}+i\check{p})_{rm}(t)  \text{,}
    \end{align}
    If we let the sum over \textit{l} be, 
    \begin{align}
        \label{tensor}
        \sum_{l}^N {\bf{V}}_{lnm} T_{lk} T_{lr}= U_{knrm} \text{,}  
    \end{align}
    which is a rank-4 tensor. This allows us to rewrite the normal mode electronic propagation as, 
    \begin{align}
        \label{electpropderivative-normalmode-tensor}
        (\dot{\check{q}} + i\dot{\check{p}})_{kn} (t) &= -i U_{knrm} 
        (\check{q}+i\check{p})_{rm}(t) \text{,}
    \end{align}
    which is very similar to Eqn.~\eqref{electpropderivative} except the potential matrix has been replaced by a tensor.

    The electronic Liouvillian, Eqn.~\eqref{l_elec_2} becomes,
    \begin{align} \label{l_elec_nonuc}
        \mathscr{L}_{\textrm{elec}, l} = \sum_{l=1}^{N} \frac{1}{\hbar} & \left[  \sum_{k,r}^N \sum_{n,m}^F\check{p}_{kn}T_{lk}{\bf{V}}_{lnm} T_{lr} \left. \frac{\partial}{\partial \check{q}_{rm}}  \right|_{\check{q}_{s \neq m, w \neq r}} \nonumber \right. \\ & \left. -  \check{q}_{kn}T_{lk}{\bf{V}}_{lnm} T_{lr} \left. \frac{\partial}{\partial \check{p}_{rm}}  \right|_{\check{p}_{s \neq m, w \neq r}} \right] \text{,}
    \end{align}
    where using Eqn.~\eqref{tensor}, such that, 
    \begin{align} \label{l_elec_nonuc-2}
        \mathscr{L}_{\textrm{elec}, l} = \frac{1}{\hbar} & \left[\sum_{k,r}^N  \sum_{n,m} ^F \check{p}_{kn} U_{knrm} \left. \frac{\partial}{\partial \check{q}_{rm}}  \right|_{\check{q}_{s \neq m, w \neq r}} \nonumber \right. \\ & \left. -  \check{q}_{kn} U_{knrm}\left. \frac{\partial}{\partial \check{p}_{rm}}  \right|_{\check{p}_{s \neq m, w \neq r}} \right]  \text{.}
    \end{align}
    Again, it is easy to show that this Liouvillian is consistent with the propagation equation in Eqn.~\eqref{electpropderivative-normalmode-tensor}.

\subsection{Conservation of the QBD}
\label{cons-qbd}
To prove conservation of the QBD, we need to show that the expectation value of $C_{AB}(t)$, where $A = \mathbb{I}$, is constant.\cite{Hele2015, Ananth2013, Hele2016} We do this by showing that the derivative is zero, noting that, 
\begin{align}
    \dot C^{[N]}_{\mathbb{I}B}(t) &= \frac{1}{Z(2\pi\hbar)^{KN}}\iint d{\bf{q}} d{\bf{p}} [e^{-\beta_N \hat{H}}]_{\overline{W}} \overrightarrow{ \mathcal{L}_\mathrm{elec}} \times [\hat{B}(t)]_W  \text{,}
\end{align}
where $\mathcal{L}_\mathrm{elec}$ is the electronic Liouvillian. We use the fact that $\overrightarrow{\mathcal{L}_\mathrm{elec}} = - \overleftarrow{\mathcal{L}_\mathrm{elec}}$, as it only consists of first-order derivatives,\cite{Hele2016} to act the Liouvillian on the zero-time $[e^{-\beta_N \hat{H}}]_{\overline{W}}$ term of the correlation function. In the following sections, we aim to prove that the Liouvillian acting on the QBD term is zero for both a single and multiple beads. 

\subsubsection{Single bead}
For 1 bead, the Liouvillian is Eqn.~\eqref{elecliou}  (where we let $\hbar=1$) and the CF is Eqn.~\eqref{CF-final-main} in the single bead limit 
such that, 
\begin{align}
    \dot{C}_{AB}(t) = - \frac{1}{Z\hbar^{2}(\pi\hbar)^{K}}& \iint  d{\bf{q}} d{\bf{p}} \ \mathcal{L}_{\textrm{elec}}  e^{-\mathcal{G}_{\textrm{tot}}/\hbar} \nonumber  \\ & \times  \textrm{Tr}\left[ e^{-\beta \bo{V}}\left( {\bf{C}} -\frac{\hbar}{2}\mathbb{I}\right)\right]  \textrm{Tr}\left[( {\bf{C}} - \hbar \mathbb{I}){\bf{B}}(t)\right] \text{,}
\end{align}
where we have used the fact that $\hat{H} = \bo{V}$ as we only have an electronic system. 

Hence, we need to prove that $\mathcal{L}_{\textrm{elec}}e^{-\mathcal{G}_{\textrm{tot}}/\hbar}=0 $ and $ \mathcal{L}_{\textrm{elec}}\textrm{Tr}\left[ e^{-\beta \bo{V}}\left( {\bf{C}} -\frac{\hbar}{2}\mathbb{I}\right)\right] =0 $ to prove conservation of the distribution. The first is easy to show utilising the fact that $\bo{V}$ is symmetric such that,
\begin{align}
    \mathcal{L}_{\textrm{elec}}e^{-\mathcal{G}_{\textrm{tot}}/\hbar} = \frac{2}{\hbar}\left(-{\bf{p}}^\textrm{T} \bo{V}  \bo{q}  + {\bf{p}}^\textrm{T} \bo{V}  \bo{q}\right)e^{-(\bo{q}^\textrm{T}\bo{q} + \bo{p}^\textrm{T}\bo{p})/\hbar} = 0 \text{.}
\end{align}
For the second part, 
\begin{align}
    \mathcal{L}_{\textrm{elec}}\textrm{Tr}\left[ e^{-\beta \bo{V}}\left( {\bf{C}} -\frac{\hbar}{2}\mathbb{I}\right)\right] 
    =& i\left( {\bf{q}} - i \bo{p}\right)^\textrm{T} \bo{V} e^{-\beta \bo{V}} \left( \bo{q} + i \bo{p}\right)  \nonumber \\ &  - i \left( {\bf{q}} - i \bo{p}\right)^\textrm{T}  e^{-\beta \bo{V}}\bo{V} \left( \bo{q} + i \bo{p}\right) \text{,}
\end{align}
where $\bo{V}$ and $e^{-\beta \bo{V}}$ commute such that these terms cancel out. 
We conclude that for a single bead in $\bo{q}$ and $\bo{p}$ the QBD is conserved.

\subsubsection{$N$ beads}
The picture is more complicated with multiple beads as the product of matrices does not possess easily identifiable commutation rules. 
For multiple beads the Liouvillian is Eqn.~\eqref{elecliou}, and we are now trying to prove that, 
\begin{subequations}
\begin{align}
    &\mathcal{L}_{\textrm{elec},l} e^{\sum_j (\bo{q}_j^\textrm{T}\bo{q}_j + \bo{p}_j^\textrm{T}\bo{p}_j ) /\hbar } = 0 \text{,}
    \\
     &\mathcal{L}_{\textrm{elec},l} \Omega    = 0 \text{,}
\end{align}
\end{subequations}
where $\mathcal{L}_{\textrm{elec},l}$ is the Liouvillian acting on the $l$-th bead and $\Omega = \textrm{Tr}\left[ \prod_j^N e^{-\beta_N \bo{V}}\left( {\bf{C}}_j -\frac{\hbar}{2}\mathbb{I}\right)\right] $ which is the Boltzmann term for $N$ beads plotted in Figure~\ref{fig:singletrajconsen}. As there are no mixed bead terms in the first condition, this follows the from the single-bead case. The second condition is more challenging to prove due to the mixed bead terms. 
We can use the cyclic trace to bring the $l$-th bead to the front,
\begin{subequations}
\begin{align}
    \Omega &= \textrm{Tr}\left[ e^{-\beta_N \bo{V}}\left( {\bf{C}}_l -\frac{\hbar}{2}\mathbb{I}\right)\prod_{k= l+1}^{k = l-1} e^{-\beta_N \bo{V}}\left( {\bf{C}}_k -\frac{\hbar}{2}\mathbb{I}\right)\right] \\
    &= \textrm{Tr}\left[ e^{-\beta_N \bo{V}}\left( {\bf{C}}_l -\frac{\hbar}{2}\mathbb{I}\right) \Gamma_l\right] \text{,}
\end{align}
\end{subequations}
where we let $\Gamma_l = \prod_{k= l+1}^{k = l-1} e^{-\beta_N \bo{V}}\left( {\bf{C}}_k -\frac{\hbar}{2}\mathbb{I}\right)$. Then,
\begin{align}
     \mathcal{L}_{\textrm{elec},l} \Omega 
     &= \sum_{l=1}^N \left( \bo{p}_l^\textrm{T} \bo{V} \overrightarrow{\nabla}_{\bo{q}_l} - \bo{q}_l^\textrm{T} \bo{V} \overrightarrow{\nabla}_{\bo{p}_l} \right) \textrm{Tr}\left[ e^{-\beta_N \bo{V}} {\bf{C}}_l \Gamma_l - \frac{\hbar}{2}e^{-\beta_N \bo{V}}\Gamma_l  \right] \text{,}
\end{align}
where $\Gamma_l$ has no dependence on the $l$-th bead that we are differentiating with respect to such that, 
\begin{subequations}
\begin{align}
     \mathcal{L}_{\textrm{elec},l} \Omega 
     =& \sum_{l=1}^N \left( \bo{p}_l^\textrm{T} \bo{V} \overrightarrow{\nabla}_{\bo{q}_l} - \bo{q}_l^\textrm{T} \bo{V} \overrightarrow{\nabla}_{\bo{p}_l} \right) \textrm{Tr}\left[ {\bf{C}}_l \Gamma_l e^{-\beta_N \bo{V}} \right] \\
       =& \sum_{l=1}^N \left( \bo{p}_l^\textrm{T} \bo{V} \overrightarrow{\nabla}_{\bo{q}_l} - \bo{q}_l^\textrm{T} \bo{V} \overrightarrow{\nabla}_{\bo{p}_l} \right) \left( {\bf{q}} - i \bo{p}\right)_l^\textrm{T} \Gamma_l  e^{-\beta_N \boldsymbol{V}}\left( {\bf{q}} + i \bo{p}\right)_l \\
     =& \sum_{l=1}^N i \left( \bo{q} - i \bo{p}\right)_l^\textrm{T} \bo{V}\Gamma_le^{-\beta_N \bo{V}}  \left( \bo{q} + i \bo{p}\right)_l \nonumber \\ &-i \left( \bo{q} - i \bo{p}\right)_l^\textrm{T}\Gamma_l e^{-\beta_N \bo{V}}\bo{V} \left( \bo{q} + i \bo{p}\right)_l \\
     =& i \sum_{l=1}^N  \textrm{Tr} \left[ \Gamma_le^{-\beta_N \bo{V}} \left( \bo{C}_l \bo{V} - \bo{V}\bo{C}_l \right) \right] \text{.}
\end{align}
\end{subequations}
If we reinsert $\hbar\mathbb
{I}/2$ terms,
\begin{align}
     \bo{C}_l \bo{V} - \bo{V}\bo{C}_l  
     &= \left( \bo{C}_l - \frac{\hbar}{2}\mathbb{I} \right) \bo{V} - \bo{V}\left(\bo{C}_l -\frac{\hbar}{2}\mathbb{I} \right) \text{,}
\end{align}
such that, 
\begin{align}
    \mathcal{L}_{\textrm{elec},l} A 
    =& i \sum_{l=1}^N \textrm{Tr} \left[ \Gamma_le^{-\beta_N \bo{V}}  \left( \bo{C}_l - \frac{\hbar}{2}\mathbb{I} \right) \bo{V} \right] \nonumber \\  &-\textrm{Tr}\left[\Gamma_le^{-\beta_N \bo{V}}  \bo{V}\left(\bo{C}_l -\frac{\hbar}{2}\mathbb{I} \right) \right]  \text{,}
\end{align}
where all terms result in a product of $\bo{C}_l -\frac{\hbar}{2}\mathbb{I}$ matrices which aids in simplifying the algebra. We refer to $e^{-\beta_N \bo{V}}  \left( \bo{C}_l - \frac{\hbar}{2}\mathbb{I} \right)$ as the Boltzmann term for the $l$-th bead.


\begin{figure}[ht]
        \centering
        \includegraphics[width=0.8\linewidth]{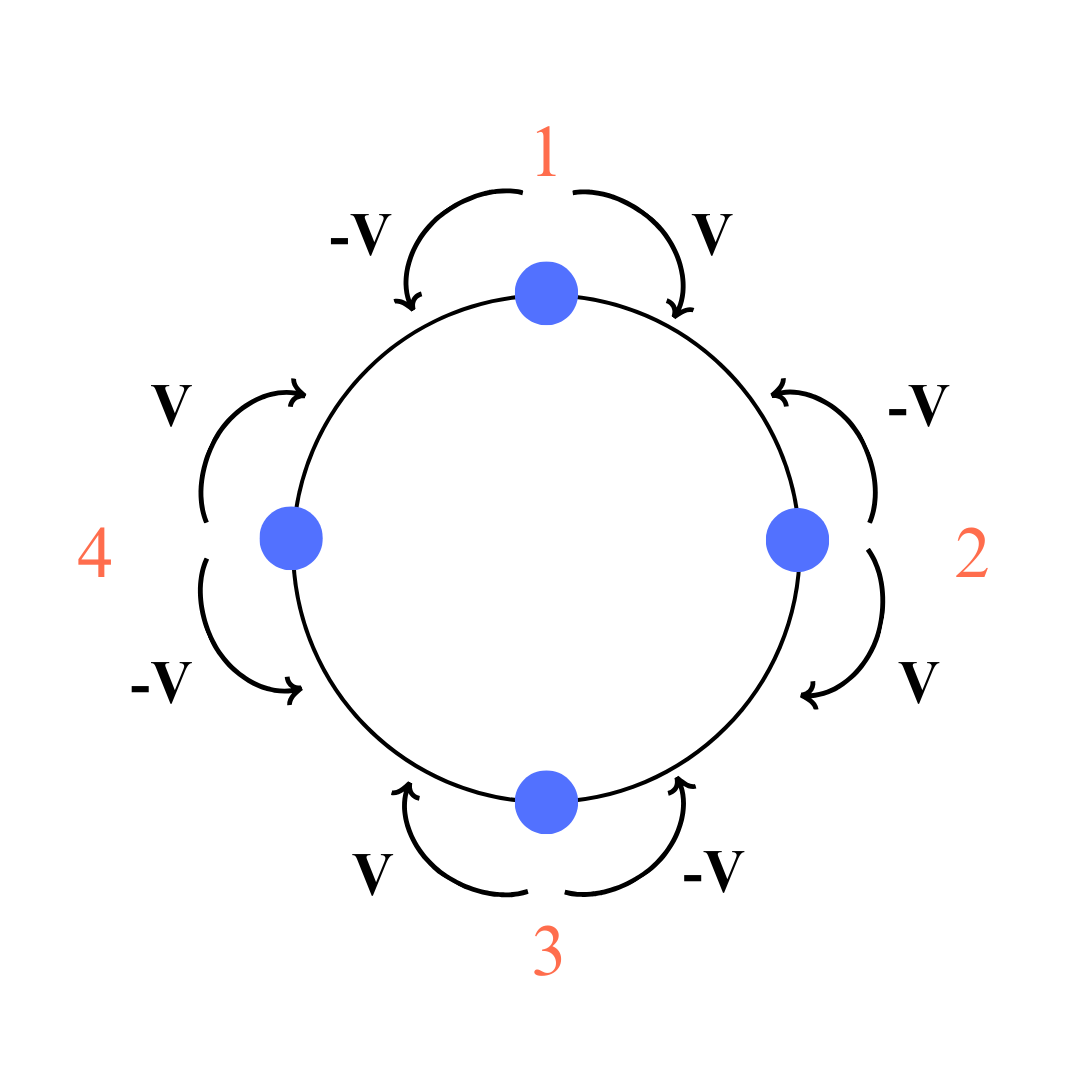}
        \caption{\label{fig:liouvillian-acting} A path integral schematic for the Liouvillian operator for 4 beads. The smooth lines are the imaginary-time evolution where the Boltzmann term is evaluated for each bead at the blue circles. We see that the Liouvillian pulls down a $\bo{V}$ to the right of the Boltzmann term and $-\bo{V}$ to the left, such that terms will cancel when the whole path integral loop is included.}
\end{figure}

We note that the Liouvillian operator results in two terms per bead which we refer to as positive and negative terms such that, for the $l$-th bead,
\begin{subequations}
\begin{align}
    \mathcal{P}_{l} 
    &=  \textrm{Tr} \left[ \Gamma_le^{-\beta_N \bo{V}}  \left( \bo{C}_l - \frac{\hbar}{2}\mathbb{I} \right) \bo{V} \right] \text{,} \\
    \mathcal{N}_{l} 
    &= - \textrm{Tr}\left[\Gamma_le^{-\beta_N \bo{V}}  \bo{V}\left(\bo{C}_l -\frac{\hbar}{2}\mathbb{I} \right) \right] \text{.}
\end{align}
\end{subequations}
The negative term can be rewritten as, 
\begin{align}
    \mathcal{N}_{l} 
    &= - \textrm{Tr}\left[\Gamma_l \bo{V} e^{-\beta_N \bo{V}} \left(\bo{C}_l -\frac{\hbar}{2}\mathbb{I} \right) \right] \text{.}
\end{align}
where we have used the fact that  $e^{-\beta_N \bo{V}} $ and $\bo{V}$ commute to reverse the order of these matrices. The two terms we obtain per bead can be seen as the Liouvillian operator `pulling' down a $\bo{V}$ to the right of the Boltzmann term for that bead and a $-\bo{V}$ to the left of the Boltzmann term for that bead. 

By using the cyclic properties of the trace and ring-polymer, we can bring the $(l+1)$-th bead to the right of the $l$-th bead such that the positive term becomes,
\begin{align}
    \mathcal{P}_l 
    &=  \textrm{Tr} \left[ Y_k e^{-\beta_N \bo{V}}  \left( \bo{C}_l - \frac{\hbar}{2}\mathbb{I} \right) \bo{V} e^{-\beta_N \bo{V}}  \left( \bo{C}_{l+1} - \frac{\hbar}{2}\mathbb{I} \right) \right] \text{,}
\end{align}
where 
\begin{align}
    Y_k = \prod_{k=l+2}^{l-1} e^{-\beta_N \bo{V}}  \left( \bo{C}_k - \frac{\hbar}{2}\mathbb{I} \right) \text{.}
\end{align}
If we do a similar trick for the $(l+1)$-th bead, but instead including the $l$-th bead to the left then the negative term for the $(l+1)$-th bead becomes,
\begin{align}
    \mathcal{N}_{l+1}
    &= - \textrm{Tr}\left[ Y_{k}  e^{-\beta_N \bo{V}}  \left(\bo{C}_{l} -\frac{\hbar}{2}\mathbb{I} \right) \bo{V}e^{-\beta_N \bo{V}}\left(\bo{C}_{l+1} -\frac{\hbar}{2}\mathbb{I} \right) \right] \text{.}
\end{align}
We now see that the positive term of the $l$-th bead results in $\bo{V}$ being in the same position in the product as for the negative term for the $(l+1)$-th bead. This results in these two terms cancelling such that when the sum over $N$ is performed,  
\begin{align}
    \mathcal{L}_{\textrm{elec},l} \Omega = \mathcal{P}_{l} + \mathcal{N}_{(l+1)} + \mathcal{P}_{l+1} + \mathcal{N}_{l+2} + \hdots + \mathcal{P}_{l-1} + \mathcal{N}_{l} =0 \text{,}
\end{align} 
as the ring-polymer is cyclic. 

We conclude that once all $N$ beads have been included, all terms cancel out and the distribution will be conserved for multiple beads in $\bo{q}$ and $\bo{p}$. This is in line with the proposed NA-Mats method conserving the QBD in the decoupled limit.\cite{Chowdhury2021} 

If all normal modes of $\bo{q}$ and $\bo{p}$ are being used then this derivation also holds true as back-transforming to beads will conserve the QBD. However, when truncating in beads or normal modes, we would expect this not to conserve the QBD as these terms will not fully cancel out. For an ensemble of trajectories, averaging may lead to conservation for both full and truncated calculations, and we leave the proof of this as future work.

\bibliography{rp}

\begin{thebibliography}{81}%
\makeatletter
\providecommand \@ifxundefined [1]{%
 \@ifx{#1\undefined}
}%
\providecommand \@ifnum [1]{%
 \ifnum #1\expandafter \@firstoftwo
 \else \expandafter \@secondoftwo
 \fi
}%
\providecommand \@ifx [1]{%
 \ifx #1\expandafter \@firstoftwo
 \else \expandafter \@secondoftwo
 \fi
}%
\providecommand \natexlab [1]{#1}%
\providecommand \enquote  [1]{``#1''}%
\providecommand \bibnamefont  [1]{#1}%
\providecommand \bibfnamefont [1]{#1}%
\providecommand \citenamefont [1]{#1}%
\providecommand \href@noop [0]{\@secondoftwo}%
\providecommand \href [0]{\begingroup \@sanitize@url \@href}%
\providecommand \@href[1]{\@@startlink{#1}\@@href}%
\providecommand \@@href[1]{\endgroup#1\@@endlink}%
\providecommand \@sanitize@url [0]{\catcode `\\12\catcode `\$12\catcode `\&12\catcode `\#12\catcode `\^12\catcode `\_12\catcode `\%12\relax}%
\providecommand \@@startlink[1]{}%
\providecommand \@@endlink[0]{}%
\providecommand \url  [0]{\begingroup\@sanitize@url \@url }%
\providecommand \@url [1]{\endgroup\@href {#1}{\urlprefix }}%
\providecommand \urlprefix  [0]{URL }%
\providecommand \Eprint [0]{\href }%
\providecommand \doibase [0]{http://dx.doi.org/}%
\providecommand \selectlanguage [0]{\@gobble}%
\providecommand \bibinfo  [0]{\@secondoftwo}%
\providecommand \bibfield  [0]{\@secondoftwo}%
\providecommand \translation [1]{[#1]}%
\providecommand \BibitemOpen [0]{}%
\providecommand \bibitemStop [0]{}%
\providecommand \bibitemNoStop [0]{.\EOS\space}%
\providecommand \EOS [0]{\spacefactor3000\relax}%
\providecommand \BibitemShut  [1]{\csname bibitem#1\endcsname}%
\let\auto@bib@innerbib\@empty
\bibitem [{\citenamefont {Polli}\ \emph {et~al.}(2010)\citenamefont {Polli}, \citenamefont {Alto{\`e}}, \citenamefont {Weingart}, \citenamefont {Spillane}, \citenamefont {Manzoni}, \citenamefont {Brida}, \citenamefont {Tomasello}, \citenamefont {Orlandi}, \citenamefont {Kukura}, \citenamefont {Mathies}, \citenamefont {Garavelli},\ and\ \citenamefont {Cerullo}}]{polliConicalIntersectionDynamics2010}%
  \BibitemOpen
  \bibfield  {author} {\bibinfo {author} {\bibfnamefont {D.}~\bibnamefont {Polli}}, \bibinfo {author} {\bibfnamefont {P.}~\bibnamefont {Alto{\`e}}}, \bibinfo {author} {\bibfnamefont {O.}~\bibnamefont {Weingart}}, \bibinfo {author} {\bibfnamefont {K.~M.}\ \bibnamefont {Spillane}}, \bibinfo {author} {\bibfnamefont {C.}~\bibnamefont {Manzoni}}, \bibinfo {author} {\bibfnamefont {D.}~\bibnamefont {Brida}}, \bibinfo {author} {\bibfnamefont {G.}~\bibnamefont {Tomasello}}, \bibinfo {author} {\bibfnamefont {G.}~\bibnamefont {Orlandi}}, \bibinfo {author} {\bibfnamefont {P.}~\bibnamefont {Kukura}}, \bibinfo {author} {\bibfnamefont {R.~A.}\ \bibnamefont {Mathies}}, \bibinfo {author} {\bibfnamefont {M.}~\bibnamefont {Garavelli}}, \ and\ \bibinfo {author} {\bibfnamefont {G.}~\bibnamefont {Cerullo}},\ }\href {\doibase 10.1038/nature09346} {\bibfield  {journal} {\bibinfo  {journal} {Nature}\ }\textbf {\bibinfo {volume} {467}},\ \bibinfo {pages} {440} (\bibinfo {year} {2010})}\BibitemShut {NoStop}%
\bibitem [{\citenamefont {Wan}\ \emph {et~al.}(1999)\citenamefont {Wan}, \citenamefont {Fiebig}, \citenamefont {Kelley}, \citenamefont {Treadway}, \citenamefont {Barton},\ and\ \citenamefont {Zewail}}]{wanFemtosecondDynamicsDNAmediated1999}%
  \BibitemOpen
  \bibfield  {author} {\bibinfo {author} {\bibfnamefont {C.}~\bibnamefont {Wan}}, \bibinfo {author} {\bibfnamefont {T.}~\bibnamefont {Fiebig}}, \bibinfo {author} {\bibfnamefont {S.~O.}\ \bibnamefont {Kelley}}, \bibinfo {author} {\bibfnamefont {C.~R.}\ \bibnamefont {Treadway}}, \bibinfo {author} {\bibfnamefont {J.~K.}\ \bibnamefont {Barton}}, \ and\ \bibinfo {author} {\bibfnamefont {A.~H.}\ \bibnamefont {Zewail}},\ }\href {\doibase 10.1073/pnas.96.11.6014} {\bibfield  {journal} {\bibinfo  {journal} {Proc. Natl. Acad. Sci.}\ }\textbf {\bibinfo {volume} {96}},\ \bibinfo {pages} {6014} (\bibinfo {year} {1999})}\BibitemShut {NoStop}%
\bibitem [{\citenamefont {{Hammes-Schiffer}}\ and\ \citenamefont {Stuchebrukhov}(2010)}]{hammes-schifferTheoryCoupledElectron2010}%
  \BibitemOpen
  \bibfield  {author} {\bibinfo {author} {\bibfnamefont {S.}~\bibnamefont {{Hammes-Schiffer}}}\ and\ \bibinfo {author} {\bibfnamefont {A.~A.}\ \bibnamefont {Stuchebrukhov}},\ }\href {\doibase 10.1021/cr1001436} {\bibfield  {journal} {\bibinfo  {journal} {Chem. Rev.}\ }\textbf {\bibinfo {volume} {110}},\ \bibinfo {pages} {6939} (\bibinfo {year} {2010})}\BibitemShut {NoStop}%
\bibitem [{\citenamefont {Marcus}\ and\ \citenamefont {Sutin}(1985)}]{marcusElectronTransfersChemistry1985}%
  \BibitemOpen
  \bibfield  {author} {\bibinfo {author} {\bibfnamefont {R.~A.}\ \bibnamefont {Marcus}}\ and\ \bibinfo {author} {\bibfnamefont {N.}~\bibnamefont {Sutin}},\ }\href {\doibase 10.1016/0304-4173(85)90014-X} {\bibfield  {journal} {\bibinfo  {journal} {Biochim. Biophys. Acta - Bioenerg.}\ }\textbf {\bibinfo {volume} {811}},\ \bibinfo {pages} {265} (\bibinfo {year} {1985})}\BibitemShut {NoStop}%
\bibitem [{\citenamefont {Zhu}\ and\ \citenamefont {Podzorov}(2015)}]{zhuChargeCarriersHybrid2015}%
  \BibitemOpen
  \bibfield  {author} {\bibinfo {author} {\bibfnamefont {X.-Y.}\ \bibnamefont {Zhu}}\ and\ \bibinfo {author} {\bibfnamefont {V.}~\bibnamefont {Podzorov}},\ }\href {\doibase 10.1021/acs.jpclett.5b02462} {\bibfield  {journal} {\bibinfo  {journal} {J. Phys. Chem. Lett.}\ }\textbf {\bibinfo {volume} {6}},\ \bibinfo {pages} {4758} (\bibinfo {year} {2015})}\BibitemShut {NoStop}%
\bibitem [{\citenamefont {Cheng}\ and\ \citenamefont {Fleming}(2009)}]{chengDynamicsLightHarvesting2009}%
  \BibitemOpen
  \bibfield  {author} {\bibinfo {author} {\bibfnamefont {Y.-C.}\ \bibnamefont {Cheng}}\ and\ \bibinfo {author} {\bibfnamefont {G.~R.}\ \bibnamefont {Fleming}},\ }\href {\doibase 10.1146/annurev.physchem.040808.090259} {\bibfield  {journal} {\bibinfo  {journal} {Annu. Rev. Phys. Chem.}\ }\textbf {\bibinfo {volume} {60}},\ \bibinfo {pages} {241} (\bibinfo {year} {2009})}\BibitemShut {NoStop}%
\bibitem [{\citenamefont {Domcke}\ and\ \citenamefont {Yarkony}(2012)}]{domckeRoleConicalIntersections2012}%
  \BibitemOpen
  \bibfield  {author} {\bibinfo {author} {\bibfnamefont {W.}~\bibnamefont {Domcke}}\ and\ \bibinfo {author} {\bibfnamefont {D.~R.}\ \bibnamefont {Yarkony}},\ }\href {\doibase 10.1146/annurev-physchem-032210-103522} {\bibfield  {journal} {\bibinfo  {journal} {Annu. Rev. Phys. Chem.}\ }\textbf {\bibinfo {volume} {63}},\ \bibinfo {pages} {325} (\bibinfo {year} {2012})}\BibitemShut {NoStop}%
\bibitem [{\citenamefont {Mukherjee}\ \emph {et~al.}(2025)\citenamefont {Mukherjee}, \citenamefont {Lassmann}, \citenamefont {Mattos}, \citenamefont {Demoulin}, \citenamefont {Curchod},\ and\ \citenamefont {Barbatti}}]{mukherjee_assessing_2025}%
  \BibitemOpen
  \bibfield  {author} {\bibinfo {author} {\bibfnamefont {S.}~\bibnamefont {Mukherjee}}, \bibinfo {author} {\bibfnamefont {Y.}~\bibnamefont {Lassmann}}, \bibinfo {author} {\bibfnamefont {R.~S.}\ \bibnamefont {Mattos}}, \bibinfo {author} {\bibfnamefont {B.}~\bibnamefont {Demoulin}}, \bibinfo {author} {\bibfnamefont {B.~F.~E.}\ \bibnamefont {Curchod}}, \ and\ \bibinfo {author} {\bibfnamefont {M.}~\bibnamefont {Barbatti}},\ }\href {\doibase 10.1021/acs.jctc.4c01349} {\bibfield  {journal} {\bibinfo  {journal} {J. Chem. Theory Comput.}\ }\textbf {\bibinfo {volume} {21}},\ \bibinfo {pages} {29} (\bibinfo {year} {2025})}\BibitemShut {NoStop}%
\bibitem [{\citenamefont {Althorpe}\ \emph {et~al.}(2016)\citenamefont {Althorpe}, \citenamefont {Angulo}, \citenamefont {Astumian}, \citenamefont {Beniwal}, \citenamefont {Bolhuis}, \citenamefont {Brand{\~a}o}, \citenamefont {Ellis}, \citenamefont {Fang}, \citenamefont {Glowacki}, \citenamefont {{Hammes-Schiffer}}, \citenamefont {Hele}, \citenamefont {J{\'o}nsson}, \citenamefont {Leli{\`e}vre}, \citenamefont {Makri}, \citenamefont {Manolopoulos}, \citenamefont {Mebel}, \citenamefont {Menzl}, \citenamefont {Miller}, \citenamefont {Parrinello}, \citenamefont {Piaggi}, \citenamefont {Pollak}, \citenamefont {Roy~Chowdhury}, \citenamefont {Sanz}, \citenamefont {Shalashilin}, \citenamefont {Sk{\'u}lason}, \citenamefont {Spezia},\ and\ \citenamefont {Taraphder}}]{Althorpe2016}%
  \BibitemOpen
  \bibfield  {author} {\bibinfo {author} {\bibfnamefont {S.}~\bibnamefont {Althorpe}}, \bibinfo {author} {\bibfnamefont {G.}~\bibnamefont {Angulo}}, \bibinfo {author} {\bibfnamefont {R.~D.}\ \bibnamefont {Astumian}}, \bibinfo {author} {\bibfnamefont {V.}~\bibnamefont {Beniwal}}, \bibinfo {author} {\bibfnamefont {P.~G.}\ \bibnamefont {Bolhuis}}, \bibinfo {author} {\bibfnamefont {J.}~\bibnamefont {Brand{\~a}o}}, \bibinfo {author} {\bibfnamefont {J.}~\bibnamefont {Ellis}}, \bibinfo {author} {\bibfnamefont {W.}~\bibnamefont {Fang}}, \bibinfo {author} {\bibfnamefont {D.~R.}\ \bibnamefont {Glowacki}}, \bibinfo {author} {\bibfnamefont {S.}~\bibnamefont {{Hammes-Schiffer}}}, \bibinfo {author} {\bibfnamefont {T.~J.~H.}\ \bibnamefont {Hele}}, \bibinfo {author} {\bibfnamefont {H.}~\bibnamefont {J{\'o}nsson}}, \bibinfo {author} {\bibfnamefont {T.}~\bibnamefont {Leli{\`e}vre}}, \bibinfo {author} {\bibfnamefont {N.}~\bibnamefont {Makri}}, \bibinfo {author} {\bibfnamefont {D.}~\bibnamefont {Manolopoulos}}, \bibinfo {author}
  {\bibfnamefont {A.~M.}\ \bibnamefont {Mebel}}, \bibinfo {author} {\bibfnamefont {G.}~\bibnamefont {Menzl}}, \bibinfo {author} {\bibfnamefont {T.~F.}\ \bibnamefont {Miller}}, \bibinfo {author} {\bibfnamefont {M.}~\bibnamefont {Parrinello}}, \bibinfo {author} {\bibfnamefont {P.~M.}\ \bibnamefont {Piaggi}}, \bibinfo {author} {\bibfnamefont {E.}~\bibnamefont {Pollak}}, \bibinfo {author} {\bibfnamefont {P.}~\bibnamefont {Roy~Chowdhury}}, \bibinfo {author} {\bibfnamefont {E.}~\bibnamefont {Sanz}}, \bibinfo {author} {\bibfnamefont {D.}~\bibnamefont {Shalashilin}}, \bibinfo {author} {\bibfnamefont {E.}~\bibnamefont {Sk{\'u}lason}}, \bibinfo {author} {\bibfnamefont {R.}~\bibnamefont {Spezia}}, \ and\ \bibinfo {author} {\bibfnamefont {S.}~\bibnamefont {Taraphder}},\ }\href {\doibase 10.1039/C6FD90076C} {\bibfield  {journal} {\bibinfo  {journal} {Faraday Discuss.}\ }\textbf {\bibinfo {volume} {195}},\ \bibinfo {pages} {671} (\bibinfo {year} {2016})}\BibitemShut {NoStop}%
\bibitem [{\citenamefont {Nelson}\ \emph {et~al.}(2020)\citenamefont {Nelson}, \citenamefont {White}, \citenamefont {Bjorgaard}, \citenamefont {Sifain}, \citenamefont {Zhang}, \citenamefont {Nebgen}, \citenamefont {{Fernandez-Alberti}}, \citenamefont {Mozyrsky}, \citenamefont {Roitberg},\ and\ \citenamefont {Tretiak}}]{nelsonNonadiabaticExcitedStateMolecular2020}%
  \BibitemOpen
  \bibfield  {author} {\bibinfo {author} {\bibfnamefont {T.~R.}\ \bibnamefont {Nelson}}, \bibinfo {author} {\bibfnamefont {A.~J.}\ \bibnamefont {White}}, \bibinfo {author} {\bibfnamefont {J.~A.}\ \bibnamefont {Bjorgaard}}, \bibinfo {author} {\bibfnamefont {A.~E.}\ \bibnamefont {Sifain}}, \bibinfo {author} {\bibfnamefont {Y.}~\bibnamefont {Zhang}}, \bibinfo {author} {\bibfnamefont {B.}~\bibnamefont {Nebgen}}, \bibinfo {author} {\bibfnamefont {S.}~\bibnamefont {{Fernandez-Alberti}}}, \bibinfo {author} {\bibfnamefont {D.}~\bibnamefont {Mozyrsky}}, \bibinfo {author} {\bibfnamefont {A.~E.}\ \bibnamefont {Roitberg}}, \ and\ \bibinfo {author} {\bibfnamefont {S.}~\bibnamefont {Tretiak}},\ }\href {\doibase 10.1021/acs.chemrev.9b00447} {\bibfield  {journal} {\bibinfo  {journal} {Chem. Rev.}\ }\textbf {\bibinfo {volume} {120}},\ \bibinfo {pages} {2215} (\bibinfo {year} {2020})}\BibitemShut {NoStop}%
\bibitem [{\citenamefont {Beck}\ \emph {et~al.}(2000)\citenamefont {Beck}, \citenamefont {J{\"a}ckle}, \citenamefont {Worth},\ and\ \citenamefont {Meyer}}]{Beck2000}%
  \BibitemOpen
  \bibfield  {author} {\bibinfo {author} {\bibfnamefont {M.~H.}\ \bibnamefont {Beck}}, \bibinfo {author} {\bibfnamefont {A.}~\bibnamefont {J{\"a}ckle}}, \bibinfo {author} {\bibfnamefont {G.~A.}\ \bibnamefont {Worth}}, \ and\ \bibinfo {author} {\bibfnamefont {H.~D.}\ \bibnamefont {Meyer}},\ }\href {\doibase 10.1016/S0370-1573(99)00047-2} {\bibfield  {journal} {\bibinfo  {journal} {Phys. Rep.}\ }\textbf {\bibinfo {volume} {324}},\ \bibinfo {pages} {1} (\bibinfo {year} {2000})}\BibitemShut {NoStop}%
\bibitem [{\citenamefont {Van~Haeften}, \citenamefont {Ash},\ and\ \citenamefont {Worth}(2023)}]{vanhaeftenPropagatingMultidimensionalDensity2023}%
  \BibitemOpen
  \bibfield  {author} {\bibinfo {author} {\bibfnamefont {A.}~\bibnamefont {Van~Haeften}}, \bibinfo {author} {\bibfnamefont {C.}~\bibnamefont {Ash}}, \ and\ \bibinfo {author} {\bibfnamefont {G.}~\bibnamefont {Worth}},\ }\href {\doibase 10.1063/5.0172956} {\bibfield  {journal} {\bibinfo  {journal} {J. Chem. Phys.}\ }\textbf {\bibinfo {volume} {159}},\ \bibinfo {pages} {194114} (\bibinfo {year} {2023})}\BibitemShut {NoStop}%
\bibitem [{\citenamefont {Wang}\ and\ \citenamefont {Thoss}(2003)}]{wangMultilayerFormulationMulticonfiguration2003}%
  \BibitemOpen
  \bibfield  {author} {\bibinfo {author} {\bibfnamefont {H.}~\bibnamefont {Wang}}\ and\ \bibinfo {author} {\bibfnamefont {M.}~\bibnamefont {Thoss}},\ }\href {\doibase 10.1063/1.1580111} {\bibfield  {journal} {\bibinfo  {journal} {J. Chem. Phys.}\ }\textbf {\bibinfo {volume} {119}},\ \bibinfo {pages} {1289} (\bibinfo {year} {2003})}\BibitemShut {NoStop}%
\bibitem [{\citenamefont {Shin}\ and\ \citenamefont {Metiu}(1996)}]{shinMultipleTimeScale1996}%
  \BibitemOpen
  \bibfield  {author} {\bibinfo {author} {\bibfnamefont {S.}~\bibnamefont {Shin}}\ and\ \bibinfo {author} {\bibfnamefont {H.}~\bibnamefont {Metiu}},\ }\href {\doibase 10.1021/jp952498a} {\bibfield  {journal} {\bibinfo  {journal} {J. Phys. Chem.}\ }\textbf {\bibinfo {volume} {100}},\ \bibinfo {pages} {7867} (\bibinfo {year} {1996})}\BibitemShut {NoStop}%
\bibitem [{\citenamefont {Stock}\ and\ \citenamefont {Thoss}(1997)}]{Stock1997}%
  \BibitemOpen
  \bibfield  {author} {\bibinfo {author} {\bibfnamefont {G.}~\bibnamefont {Stock}}\ and\ \bibinfo {author} {\bibfnamefont {M.}~\bibnamefont {Thoss}},\ }\href {\doibase 10.1103/PhysRevLett.78.578} {\bibfield  {journal} {\bibinfo  {journal} {Phys. Rev. Lett.}\ }\textbf {\bibinfo {volume} {78}},\ \bibinfo {pages} {578} (\bibinfo {year} {1997})}\BibitemShut {NoStop}%
\bibitem [{\citenamefont {Miller}(1970)}]{Miller1970}%
  \BibitemOpen
  \bibfield  {author} {\bibinfo {author} {\bibfnamefont {W.~H.}\ \bibnamefont {Miller}},\ }\href {\doibase 10.1063/1.1674535} {\bibfield  {journal} {\bibinfo  {journal} {J. Chem. Phys.}\ }\textbf {\bibinfo {volume} {53}},\ \bibinfo {pages} {3578} (\bibinfo {year} {1970})}\BibitemShut {NoStop}%
\bibitem [{\citenamefont {Wang}, \citenamefont {Sun},\ and\ \citenamefont {Miller}(1998)}]{Wang1998}%
  \BibitemOpen
  \bibfield  {author} {\bibinfo {author} {\bibfnamefont {H.}~\bibnamefont {Wang}}, \bibinfo {author} {\bibfnamefont {X.}~\bibnamefont {Sun}}, \ and\ \bibinfo {author} {\bibfnamefont {W.~H.}\ \bibnamefont {Miller}},\ }\href {\doibase 10.1063/1.476447} {\bibfield  {journal} {\bibinfo  {journal} {J. Chem. Phys.}\ }\textbf {\bibinfo {volume} {108}},\ \bibinfo {pages} {9726} (\bibinfo {year} {1998})}\BibitemShut {NoStop}%
\bibitem [{\citenamefont {Sun}, \citenamefont {Wang},\ and\ \citenamefont {Miller}(1998{\natexlab{a}})}]{Sun1998}%
  \BibitemOpen
  \bibfield  {author} {\bibinfo {author} {\bibfnamefont {X.}~\bibnamefont {Sun}}, \bibinfo {author} {\bibfnamefont {H.}~\bibnamefont {Wang}}, \ and\ \bibinfo {author} {\bibfnamefont {W.~H.}\ \bibnamefont {Miller}},\ }\href {\doibase 10.1063/1.477025} {\bibfield  {journal} {\bibinfo  {journal} {J. Chem. Phys.}\ }\textbf {\bibinfo {volume} {109}},\ \bibinfo {pages} {4190} (\bibinfo {year} {1998}{\natexlab{a}})}\BibitemShut {NoStop}%
\bibitem [{\citenamefont {Sun}, \citenamefont {Wang},\ and\ \citenamefont {Miller}(1998{\natexlab{b}})}]{Sun1998b}%
  \BibitemOpen
  \bibfield  {author} {\bibinfo {author} {\bibfnamefont {X.}~\bibnamefont {Sun}}, \bibinfo {author} {\bibfnamefont {H.}~\bibnamefont {Wang}}, \ and\ \bibinfo {author} {\bibfnamefont {W.~H.}\ \bibnamefont {Miller}},\ }\href {\doibase 10.1063/1.477389} {\bibfield  {journal} {\bibinfo  {journal} {J. Chem. Phys.}\ }\textbf {\bibinfo {volume} {109}},\ \bibinfo {pages} {7064} (\bibinfo {year} {1998}{\natexlab{b}})}\BibitemShut {NoStop}%
\bibitem [{\citenamefont {Church}\ \emph {et~al.}(2018)\citenamefont {Church}, \citenamefont {Hele}, \citenamefont {Ezra},\ and\ \citenamefont {Ananth}}]{Church2018}%
  \BibitemOpen
  \bibfield  {author} {\bibinfo {author} {\bibfnamefont {M.~S.}\ \bibnamefont {Church}}, \bibinfo {author} {\bibfnamefont {T.~J.~H.}\ \bibnamefont {Hele}}, \bibinfo {author} {\bibfnamefont {G.~S.}\ \bibnamefont {Ezra}}, \ and\ \bibinfo {author} {\bibfnamefont {N.}~\bibnamefont {Ananth}},\ }\href {\doibase 10.1063/1.5005557} {\bibfield  {journal} {\bibinfo  {journal} {J. Chem. Phys.}\ }\textbf {\bibinfo {volume} {148}},\ \bibinfo {pages} {102326} (\bibinfo {year} {2018})}\BibitemShut {NoStop}%
\bibitem [{\citenamefont {Ananth}, \citenamefont {Venkataraman},\ and\ \citenamefont {Miller}(2007)}]{Ananth2007}%
  \BibitemOpen
  \bibfield  {author} {\bibinfo {author} {\bibfnamefont {N.}~\bibnamefont {Ananth}}, \bibinfo {author} {\bibfnamefont {C.}~\bibnamefont {Venkataraman}}, \ and\ \bibinfo {author} {\bibfnamefont {W.~H.}\ \bibnamefont {Miller}},\ }\href {\doibase 10.1063/1.2759932} {\bibfield  {journal} {\bibinfo  {journal} {J. Chem. Phys.}\ }\textbf {\bibinfo {volume} {127}},\ \bibinfo {pages} {084114} (\bibinfo {year} {2007})}\BibitemShut {NoStop}%
\bibitem [{\citenamefont {Richardson}\ and\ \citenamefont {Thoss}(2013)}]{Richardson2013}%
  \BibitemOpen
  \bibfield  {author} {\bibinfo {author} {\bibfnamefont {J.~O.}\ \bibnamefont {Richardson}}\ and\ \bibinfo {author} {\bibfnamefont {M.}~\bibnamefont {Thoss}},\ }\href {\doibase 10.1063/1.4816124} {\bibfield  {journal} {\bibinfo  {journal} {J. Chem. Phys.}\ }\textbf {\bibinfo {volume} {139}},\ \bibinfo {pages} {031102} (\bibinfo {year} {2013})}\BibitemShut {NoStop}%
\bibitem [{\citenamefont {Ananth}(2013)}]{Ananth2013}%
  \BibitemOpen
  \bibfield  {author} {\bibinfo {author} {\bibfnamefont {N.}~\bibnamefont {Ananth}},\ }\href {\doibase 10.1063/1.4821590} {\bibfield  {journal} {\bibinfo  {journal} {J. Chem. Phys.}\ }\textbf {\bibinfo {volume} {139}},\ \bibinfo {pages} {124102} (\bibinfo {year} {2013})}\BibitemShut {NoStop}%
\bibitem [{\citenamefont {Hele}(2011)}]{Hele2011}%
  \BibitemOpen
  \bibfield  {author} {\bibinfo {author} {\bibfnamefont {T.~J.~H.}\ \bibnamefont {Hele}},\ }\emph {\bibinfo {title} {An {{Electronically Non-Adiabatic Generalization}} of {{Ring Polymer Molecular Dynamics}}}},\ \href {\doibase 10.48550/arxiv.1308.3950} {Ph.D. thesis},\ \bibinfo  {school} {University of Oxford} (\bibinfo {year} {2011})\BibitemShut {NoStop}%
\bibitem [{\citenamefont {Hele}\ \emph {et~al.}(2015{\natexlab{a}})\citenamefont {Hele}, \citenamefont {Willatt}, \citenamefont {Muolo},\ and\ \citenamefont {Althorpe}}]{Hele2015}%
  \BibitemOpen
  \bibfield  {author} {\bibinfo {author} {\bibfnamefont {T.~J.~H.}\ \bibnamefont {Hele}}, \bibinfo {author} {\bibfnamefont {M.~J.}\ \bibnamefont {Willatt}}, \bibinfo {author} {\bibfnamefont {A.}~\bibnamefont {Muolo}}, \ and\ \bibinfo {author} {\bibfnamefont {S.~C.}\ \bibnamefont {Althorpe}},\ }\href {\doibase 10.1063/1.4916311} {\bibfield  {journal} {\bibinfo  {journal} {J. Chem. Phys.}\ }\textbf {\bibinfo {volume} {142}},\ \bibinfo {pages} {134103} (\bibinfo {year} {2015}{\natexlab{a}})}\BibitemShut {NoStop}%
\bibitem [{\citenamefont {Cao}\ and\ \citenamefont {Voth}(1993)}]{Cao1993}%
  \BibitemOpen
  \bibfield  {author} {\bibinfo {author} {\bibfnamefont {J.}~\bibnamefont {Cao}}\ and\ \bibinfo {author} {\bibfnamefont {G.~A.}\ \bibnamefont {Voth}},\ }\href {\doibase 10.1063/1.465512} {\bibfield  {journal} {\bibinfo  {journal} {J. Chem. Phys.}\ }\textbf {\bibinfo {volume} {99}},\ \bibinfo {pages} {10070} (\bibinfo {year} {1993})}\BibitemShut {NoStop}%
\bibitem [{\citenamefont {Cao}\ and\ \citenamefont {Voth}(1994{\natexlab{a}})}]{Cao1994}%
  \BibitemOpen
  \bibfield  {author} {\bibinfo {author} {\bibfnamefont {J.}~\bibnamefont {Cao}}\ and\ \bibinfo {author} {\bibfnamefont {G.~A.}\ \bibnamefont {Voth}},\ }\href {\doibase 10.1063/1.467176} {\bibfield  {journal} {\bibinfo  {journal} {J. Chem. Phys.}\ }\textbf {\bibinfo {volume} {100}},\ \bibinfo {pages} {5106} (\bibinfo {year} {1994}{\natexlab{a}})}\BibitemShut {NoStop}%
\bibitem [{\citenamefont {Cao}\ and\ \citenamefont {Voth}(1994{\natexlab{b}})}]{Cao1994a}%
  \BibitemOpen
  \bibfield  {author} {\bibinfo {author} {\bibfnamefont {J.}~\bibnamefont {Cao}}\ and\ \bibinfo {author} {\bibfnamefont {G.~A.}\ \bibnamefont {Voth}},\ }\href {\doibase 10.1063/1.468503} {\bibfield  {journal} {\bibinfo  {journal} {J. Chem. Phys.}\ }\textbf {\bibinfo {volume} {101}},\ \bibinfo {pages} {6157} (\bibinfo {year} {1994}{\natexlab{b}})}\BibitemShut {NoStop}%
\bibitem [{\citenamefont {Cao}\ and\ \citenamefont {Voth}(1994{\natexlab{c}})}]{Cao1994b}%
  \BibitemOpen
  \bibfield  {author} {\bibinfo {author} {\bibfnamefont {J.}~\bibnamefont {Cao}}\ and\ \bibinfo {author} {\bibfnamefont {G.~A.}\ \bibnamefont {Voth}},\ }\href {\doibase 10.1063/1.467175} {\bibfield  {journal} {\bibinfo  {journal} {J. Chem. Phys.}\ }\textbf {\bibinfo {volume} {100}},\ \bibinfo {pages} {5093} (\bibinfo {year} {1994}{\natexlab{c}})}\BibitemShut {NoStop}%
\bibitem [{\citenamefont {Rossi}, \citenamefont {Ceriotti},\ and\ \citenamefont {Manolopoulos}(2014)}]{Rossi2014}%
  \BibitemOpen
  \bibfield  {author} {\bibinfo {author} {\bibfnamefont {M.}~\bibnamefont {Rossi}}, \bibinfo {author} {\bibfnamefont {M.}~\bibnamefont {Ceriotti}}, \ and\ \bibinfo {author} {\bibfnamefont {D.~E.}\ \bibnamefont {Manolopoulos}},\ }\href {\doibase 10.1063/1.4883861} {\bibfield  {journal} {\bibinfo  {journal} {J. Chem. Phys.}\ }\textbf {\bibinfo {volume} {140}},\ \bibinfo {pages} {234116} (\bibinfo {year} {2014})}\BibitemShut {NoStop}%
\bibitem [{\citenamefont {Kapral}(2016)}]{Kapral2016}%
  \BibitemOpen
  \bibfield  {author} {\bibinfo {author} {\bibfnamefont {R.}~\bibnamefont {Kapral}},\ }\href {\doibase 10.1016/j.chemphys.2016.05.016} {\bibfield  {journal} {\bibinfo  {journal} {Chem. Phys.}\ }\textbf {\bibinfo {volume} {481}},\ \bibinfo {pages} {77} (\bibinfo {year} {2016})}\BibitemShut {NoStop}%
\bibitem [{\citenamefont {Tully}\ and\ \citenamefont {Preston}(1971)}]{Tully1971}%
  \BibitemOpen
  \bibfield  {author} {\bibinfo {author} {\bibfnamefont {J.~C.}\ \bibnamefont {Tully}}\ and\ \bibinfo {author} {\bibfnamefont {R.~K.}\ \bibnamefont {Preston}},\ }\href {\doibase 10.1063/1.1675788} {\bibfield  {journal} {\bibinfo  {journal} {J. Chem. Phys.}\ }\textbf {\bibinfo {volume} {55}},\ \bibinfo {pages} {562} (\bibinfo {year} {1971})}\BibitemShut {NoStop}%
\bibitem [{\citenamefont {Tully}(2012)}]{tullyPerspectiveNonadiabaticDynamics2012}%
  \BibitemOpen
  \bibfield  {author} {\bibinfo {author} {\bibfnamefont {J.~C.}\ \bibnamefont {Tully}},\ }\href {\doibase 10.1063/1.4757762} {\bibfield  {journal} {\bibinfo  {journal} {J. Chem. Phys.}\ }\textbf {\bibinfo {volume} {137}},\ \bibinfo {pages} {22A301} (\bibinfo {year} {2012})}\BibitemShut {NoStop}%
\bibitem [{\citenamefont {Shakib}\ and\ \citenamefont {Huo}(2017)}]{Shakib2017}%
  \BibitemOpen
  \bibfield  {author} {\bibinfo {author} {\bibfnamefont {F.~A.}\ \bibnamefont {Shakib}}\ and\ \bibinfo {author} {\bibfnamefont {P.}~\bibnamefont {Huo}},\ }\href {\doibase 10.1021/acs.jpclett.7b01343} {\bibfield  {journal} {\bibinfo  {journal} {J. Phys. Chem. Lett.}\ }\textbf {\bibinfo {volume} {8}},\ \bibinfo {pages} {3073} (\bibinfo {year} {2017})}\BibitemShut {NoStop}%
\bibitem [{\citenamefont {Shalashilin}(2011)}]{Shalashilin2011}%
  \BibitemOpen
  \bibfield  {author} {\bibinfo {author} {\bibfnamefont {D.~V.}\ \bibnamefont {Shalashilin}},\ }\href {\doibase 10.1039/c1fd00034a} {\bibfield  {journal} {\bibinfo  {journal} {Faraday Discuss.}\ }\textbf {\bibinfo {volume} {153}},\ \bibinfo {pages} {105} (\bibinfo {year} {2011})}\BibitemShut {NoStop}%
\bibitem [{\citenamefont {Zimmermann}\ and\ \citenamefont {Van{\'i}{\v c}ek}(2014)}]{Zimmermann2014}%
  \BibitemOpen
  \bibfield  {author} {\bibinfo {author} {\bibfnamefont {T.}~\bibnamefont {Zimmermann}}\ and\ \bibinfo {author} {\bibfnamefont {J.}~\bibnamefont {Van{\'i}{\v c}ek}},\ }\href {\doibase 10.1063/1.4896735} {\bibfield  {journal} {\bibinfo  {journal} {J. Chem. Phys.}\ }\textbf {\bibinfo {volume} {141}},\ \bibinfo {pages} {134102} (\bibinfo {year} {2014})}\BibitemShut {NoStop}%
\bibitem [{\citenamefont {Meyer}\ and\ \citenamefont {Miller}(1979)}]{Meyer1979}%
  \BibitemOpen
  \bibfield  {author} {\bibinfo {author} {\bibfnamefont {H.-D.}\ \bibnamefont {Meyer}}\ and\ \bibinfo {author} {\bibfnamefont {W.~H.}\ \bibnamefont {Miller}},\ }\href {\doibase 10.1063/1.437910} {\bibfield  {journal} {\bibinfo  {journal} {J. Chem. Phys.}\ }\textbf {\bibinfo {volume} {70}},\ \bibinfo {pages} {3214} (\bibinfo {year} {1979})}\BibitemShut {NoStop}%
\bibitem [{\citenamefont {Stock}\ and\ \citenamefont {Thoss}(2005)}]{Stock2005}%
  \BibitemOpen
  \bibfield  {author} {\bibinfo {author} {\bibfnamefont {G.}~\bibnamefont {Stock}}\ and\ \bibinfo {author} {\bibfnamefont {M.}~\bibnamefont {Thoss}},\ }in\ \href {\doibase 10.1002/0471739464.ch5} {\emph {\bibinfo {booktitle} {Advances in {{Chemical Physics}}}}},\ Vol.\ \bibinfo {volume} {131},\ \bibinfo {editor} {edited by\ \bibinfo {editor} {\bibfnamefont {S.~A.}\ \bibnamefont {Rice}}}\ (\bibinfo  {publisher} {John Wiley \& Sons},\ \bibinfo {year} {2005})\ pp.\ \bibinfo {pages} {243--375}\BibitemShut {NoStop}%
\bibitem [{\citenamefont {Runeson}\ and\ \citenamefont {Richardson}(2019)}]{Runeson2019}%
  \BibitemOpen
  \bibfield  {author} {\bibinfo {author} {\bibfnamefont {J.~E.}\ \bibnamefont {Runeson}}\ and\ \bibinfo {author} {\bibfnamefont {J.~O.}\ \bibnamefont {Richardson}},\ }\href {\doibase 10.1063/1.5100506} {\bibfield  {journal} {\bibinfo  {journal} {J. Chem. Phys.}\ }\textbf {\bibinfo {volume} {151}},\ \bibinfo {pages} {044119} (\bibinfo {year} {2019})}\BibitemShut {NoStop}%
\bibitem [{\citenamefont {Mannouch}\ and\ \citenamefont {Richardson}(2023)}]{MASH}%
  \BibitemOpen
  \bibfield  {author} {\bibinfo {author} {\bibfnamefont {J.~R.}\ \bibnamefont {Mannouch}}\ and\ \bibinfo {author} {\bibfnamefont {J.~O.}\ \bibnamefont {Richardson}},\ }\href {\doibase 10.1063/5.0139734} {\bibfield  {journal} {\bibinfo  {journal} {J. Chem. Phys.}\ }\textbf {\bibinfo {volume} {158}},\ \bibinfo {pages} {104111} (\bibinfo {year} {2023})}\BibitemShut {NoStop}%
\bibitem [{\citenamefont {Runeson}\ and\ \citenamefont {Richardson}(2020)}]{Runeson2020}%
  \BibitemOpen
  \bibfield  {author} {\bibinfo {author} {\bibfnamefont {J.~E.}\ \bibnamefont {Runeson}}\ and\ \bibinfo {author} {\bibfnamefont {J.~O.}\ \bibnamefont {Richardson}},\ }\href {\doibase 10.1063/1.5143412} {\bibfield  {journal} {\bibinfo  {journal} {J. Chem. Phys.}\ }\textbf {\bibinfo {volume} {152}},\ \bibinfo {pages} {084110} (\bibinfo {year} {2020})}\BibitemShut {NoStop}%
\bibitem [{\citenamefont {Runeson}\ and\ \citenamefont {Richardson}(2021)}]{Runeson2021}%
  \BibitemOpen
  \bibfield  {author} {\bibinfo {author} {\bibfnamefont {J.~E.}\ \bibnamefont {Runeson}}\ and\ \bibinfo {author} {\bibfnamefont {J.~O.}\ \bibnamefont {Richardson}},\ }\href {\doibase 10.1103/PhysRevLett.127.250403} {\bibfield  {journal} {\bibinfo  {journal} {Phys. Rev. Lett.}\ }\textbf {\bibinfo {volume} {127}},\ \bibinfo {pages} {250403} (\bibinfo {year} {2021})}\BibitemShut {NoStop}%
\bibitem [{\citenamefont {Amati}, \citenamefont {Runeson},\ and\ \citenamefont {Richardson}(2023)}]{Amati2023}%
  \BibitemOpen
  \bibfield  {author} {\bibinfo {author} {\bibfnamefont {G.}~\bibnamefont {Amati}}, \bibinfo {author} {\bibfnamefont {J.~E.}\ \bibnamefont {Runeson}}, \ and\ \bibinfo {author} {\bibfnamefont {J.~O.}\ \bibnamefont {Richardson}},\ }\href {\doibase 10.1063/5.0137828} {\bibfield  {journal} {\bibinfo  {journal} {J. Chem. Phys.}\ }\textbf {\bibinfo {volume} {158}},\ \bibinfo {pages} {064113} (\bibinfo {year} {2023})}\BibitemShut {NoStop}%
\bibitem [{\citenamefont {Mannouch}\ and\ \citenamefont {Richardson}(2020)}]{Mannouch2020}%
  \BibitemOpen
  \bibfield  {author} {\bibinfo {author} {\bibfnamefont {J.~R.}\ \bibnamefont {Mannouch}}\ and\ \bibinfo {author} {\bibfnamefont {J.~O.}\ \bibnamefont {Richardson}},\ }\href {\doibase 10.1063/5.0031168} {\bibfield  {journal} {\bibinfo  {journal} {J. Chem. Phys.}\ }\textbf {\bibinfo {volume} {153}},\ \bibinfo {pages} {194109} (\bibinfo {year} {2020})}\BibitemShut {NoStop}%
\bibitem [{\citenamefont {Richardson}, \citenamefont {Lawrence},\ and\ \citenamefont {Mannouch}(2025)}]{richardsonNonadiabaticDynamicsMapping2025}%
  \BibitemOpen
  \bibfield  {author} {\bibinfo {author} {\bibfnamefont {J.~O.}\ \bibnamefont {Richardson}}, \bibinfo {author} {\bibfnamefont {J.~E.}\ \bibnamefont {Lawrence}}, \ and\ \bibinfo {author} {\bibfnamefont {J.~R.}\ \bibnamefont {Mannouch}},\ }\href {\doibase 10.1146/annurev-physchem-082423-120631} {\bibfield  {journal} {\bibinfo  {journal} {Annu. Rev. Phys. Chem.}\ }\textbf {\bibinfo {volume} {76}},\ \bibinfo {pages} {663} (\bibinfo {year} {2025})}\BibitemShut {NoStop}%
\bibitem [{\citenamefont {Runeson}, \citenamefont {Fay},\ and\ \citenamefont {Manolopoulos}(2024)}]{runeson_exciton_2024}%
  \BibitemOpen
  \bibfield  {author} {\bibinfo {author} {\bibfnamefont {J.~E.}\ \bibnamefont {Runeson}}, \bibinfo {author} {\bibfnamefont {T.~P.}\ \bibnamefont {Fay}}, \ and\ \bibinfo {author} {\bibfnamefont {D.~E.}\ \bibnamefont {Manolopoulos}},\ }\href {\doibase 10.1039/D3CP05926J} {\bibfield  {journal} {\bibinfo  {journal} {Phys. Chem. Chem. Phys.}\ }\textbf {\bibinfo {volume} {26}},\ \bibinfo {pages} {4929} (\bibinfo {year} {2024})}\BibitemShut {NoStop}%
\bibitem [{\citenamefont {Lawrence}, \citenamefont {Mannouch},\ and\ \citenamefont {Richardson}(2024)}]{lawrence_size-consistent_2024}%
  \BibitemOpen
  \bibfield  {author} {\bibinfo {author} {\bibfnamefont {J.~E.}\ \bibnamefont {Lawrence}}, \bibinfo {author} {\bibfnamefont {J.~R.}\ \bibnamefont {Mannouch}}, \ and\ \bibinfo {author} {\bibfnamefont {J.~O.}\ \bibnamefont {Richardson}},\ }\href {\doibase 10.1063/5.0208575} {\bibfield  {journal} {\bibinfo  {journal} {J. Chem. Phys.}\ }\textbf {\bibinfo {volume} {160}},\ \bibinfo {pages} {244112} (\bibinfo {year} {2024})}\BibitemShut {NoStop}%
\bibitem [{\citenamefont {Geuther}, \citenamefont {Asnaashari},\ and\ \citenamefont {Richardson}(2025)}]{geutherTimeReversibleImplementationMASH2025}%
  \BibitemOpen
  \bibfield  {author} {\bibinfo {author} {\bibfnamefont {J.~A.}\ \bibnamefont {Geuther}}, \bibinfo {author} {\bibfnamefont {K.}~\bibnamefont {Asnaashari}}, \ and\ \bibinfo {author} {\bibfnamefont {J.~O.}\ \bibnamefont {Richardson}},\ }\href {\doibase 10.1021/acs.jctc.4c01684} {\bibfield  {journal} {\bibinfo  {journal} {J. Chem. Theory Comput.}\ }\textbf {\bibinfo {volume} {21}},\ \bibinfo {pages} {2179} (\bibinfo {year} {2025})}\BibitemShut {NoStop}%
\bibitem [{\citenamefont {Cook}\ \emph {et~al.}(2023)\citenamefont {Cook}, \citenamefont {Runeson}, \citenamefont {Richardson},\ and\ \citenamefont {Hele}}]{Cook2023}%
  \BibitemOpen
  \bibfield  {author} {\bibinfo {author} {\bibfnamefont {L.~E.}\ \bibnamefont {Cook}}, \bibinfo {author} {\bibfnamefont {J.~E.}\ \bibnamefont {Runeson}}, \bibinfo {author} {\bibfnamefont {J.~O.}\ \bibnamefont {Richardson}}, \ and\ \bibinfo {author} {\bibfnamefont {T.~J.~H.}\ \bibnamefont {Hele}},\ }\href {\doibase 10.1021/acs.jctc.3c00709} {\bibfield  {journal} {\bibinfo  {journal} {J. Chem. Theory Comput.}\ }\textbf {\bibinfo {volume} {19}},\ \bibinfo {pages} {6109} (\bibinfo {year} {2023})}\BibitemShut {NoStop}%
\bibitem [{\citenamefont {Willatt}(2017)}]{Willatt2017}%
  \BibitemOpen
  \bibfield  {author} {\bibinfo {author} {\bibfnamefont {M.~J.}\ \bibnamefont {Willatt}},\ }\emph {\bibinfo {title} {Matsubara Dynamics and Its Practical Implementation}},\ \href {\doibase 10.17863/CAM.13644} {Ph.D. thesis},\ \bibinfo  {school} {University of Cambridge} (\bibinfo {year} {2017})\BibitemShut {NoStop}%
\bibitem [{\citenamefont {Hele}\ and\ \citenamefont {Althorpe}(2013{\natexlab{a}})}]{Hele2013}%
  \BibitemOpen
  \bibfield  {author} {\bibinfo {author} {\bibfnamefont {T.~J.~H.}\ \bibnamefont {Hele}}\ and\ \bibinfo {author} {\bibfnamefont {S.~C.}\ \bibnamefont {Althorpe}},\ }\href {\doibase 10.1063/1.4792697} {\bibfield  {journal} {\bibinfo  {journal} {J. Chem. Phys.}\ }\textbf {\bibinfo {volume} {138}},\ \bibinfo {pages} {084108} (\bibinfo {year} {2013}{\natexlab{a}})}\BibitemShut {NoStop}%
\bibitem [{\citenamefont {Hele}\ and\ \citenamefont {Althorpe}(2013{\natexlab{b}})}]{Hele2013b}%
  \BibitemOpen
  \bibfield  {author} {\bibinfo {author} {\bibfnamefont {T.~J.~H.}\ \bibnamefont {Hele}}\ and\ \bibinfo {author} {\bibfnamefont {S.~C.}\ \bibnamefont {Althorpe}},\ }\href {\doibase 10.1063/1.4819077} {\bibfield  {journal} {\bibinfo  {journal} {J. Chem. Phys.}\ }\textbf {\bibinfo {volume} {139}},\ \bibinfo {pages} {084116} (\bibinfo {year} {2013}{\natexlab{b}})}\BibitemShut {NoStop}%
\bibitem [{\citenamefont {Althorpe}\ and\ \citenamefont {Hele}(2013)}]{Althorpe2013}%
  \BibitemOpen
  \bibfield  {author} {\bibinfo {author} {\bibfnamefont {S.~C.}\ \bibnamefont {Althorpe}}\ and\ \bibinfo {author} {\bibfnamefont {T.~J.~H.}\ \bibnamefont {Hele}},\ }\href {\doibase 10.1063/1.4819076} {\bibfield  {journal} {\bibinfo  {journal} {J. Chem. Phys.}\ }\textbf {\bibinfo {volume} {139}},\ \bibinfo {pages} {084115} (\bibinfo {year} {2013})}\BibitemShut {NoStop}%
\bibitem [{\citenamefont {Hele}(2014)}]{Hele2014}%
  \BibitemOpen
  \bibfield  {author} {\bibinfo {author} {\bibfnamefont {T.~J.~H.}\ \bibnamefont {Hele}},\ }\emph {\bibinfo {title} {Quantum {{Transition-State Theory}}}},\ \href@noop {} {Ph.D. thesis},\ \bibinfo  {school} {University of Cambridge} (\bibinfo {year} {2014})\BibitemShut {NoStop}%
\bibitem [{\citenamefont {Hele}\ and\ \citenamefont {Althorpe}(2016)}]{hele_alternative_2016}%
  \BibitemOpen
  \bibfield  {author} {\bibinfo {author} {\bibfnamefont {T.~J.~H.}\ \bibnamefont {Hele}}\ and\ \bibinfo {author} {\bibfnamefont {S.~C.}\ \bibnamefont {Althorpe}},\ }\href {\doibase 10.1063/1.4947589} {\bibfield  {journal} {\bibinfo  {journal} {J. Chem. Phys.}\ }\textbf {\bibinfo {volume} {144}},\ \bibinfo {pages} {174107} (\bibinfo {year} {2016})}\BibitemShut {NoStop}%
\bibitem [{\citenamefont {Chowdhury}\ and\ \citenamefont {Huo}(2021)}]{Chowdhury2021}%
  \BibitemOpen
  \bibfield  {author} {\bibinfo {author} {\bibfnamefont {S.~N.}\ \bibnamefont {Chowdhury}}\ and\ \bibinfo {author} {\bibfnamefont {P.}~\bibnamefont {Huo}},\ }\href {\doibase 10.1063/5.0042136} {\bibfield  {journal} {\bibinfo  {journal} {J. Chem. Phys.}\ }\textbf {\bibinfo {volume} {154}},\ \bibinfo {pages} {124124} (\bibinfo {year} {2021})}\BibitemShut {NoStop}%
\bibitem [{\citenamefont {Richardson}\ \emph {et~al.}(2017)\citenamefont {Richardson}, \citenamefont {Meyer}, \citenamefont {Pleinert},\ and\ \citenamefont {Thoss}}]{Richardson2017}%
  \BibitemOpen
  \bibfield  {author} {\bibinfo {author} {\bibfnamefont {J.~O.}\ \bibnamefont {Richardson}}, \bibinfo {author} {\bibfnamefont {P.}~\bibnamefont {Meyer}}, \bibinfo {author} {\bibfnamefont {M.-O.}\ \bibnamefont {Pleinert}}, \ and\ \bibinfo {author} {\bibfnamefont {M.}~\bibnamefont {Thoss}},\ }\href {\doibase 10.1016/j.chemphys.2016.09.036} {\bibfield  {journal} {\bibinfo  {journal} {Chem. Phys.}\ }\textbf {\bibinfo {volume} {482}},\ \bibinfo {pages} {124} (\bibinfo {year} {2017})}\BibitemShut {NoStop}%
\bibitem [{\citenamefont {Craig}\ and\ \citenamefont {Manolopoulos}(2004)}]{Craig2004}%
  \BibitemOpen
  \bibfield  {author} {\bibinfo {author} {\bibfnamefont {I.~R.}\ \bibnamefont {Craig}}\ and\ \bibinfo {author} {\bibfnamefont {D.~E.}\ \bibnamefont {Manolopoulos}},\ }\href {\doibase 10.1063/1.1777575} {\bibfield  {journal} {\bibinfo  {journal} {J. Chem. Phys.}\ }\textbf {\bibinfo {volume} {121}},\ \bibinfo {pages} {3368} (\bibinfo {year} {2004})}\BibitemShut {NoStop}%
\bibitem [{\citenamefont {Ceriotti}\ \emph {et~al.}(2010)\citenamefont {Ceriotti}, \citenamefont {Parrinello}, \citenamefont {Markland},\ and\ \citenamefont {Manolopoulos}}]{ceriottiEfficientStochasticThermostatting2010}%
  \BibitemOpen
  \bibfield  {author} {\bibinfo {author} {\bibfnamefont {M.}~\bibnamefont {Ceriotti}}, \bibinfo {author} {\bibfnamefont {M.}~\bibnamefont {Parrinello}}, \bibinfo {author} {\bibfnamefont {T.~E.}\ \bibnamefont {Markland}}, \ and\ \bibinfo {author} {\bibfnamefont {D.~E.}\ \bibnamefont {Manolopoulos}},\ }\href {\doibase 10.1063/1.3489925} {\bibfield  {journal} {\bibinfo  {journal} {J. Chem. Phys.}\ }\textbf {\bibinfo {volume} {133}},\ \bibinfo {pages} {124104} (\bibinfo {year} {2010})}\BibitemShut {NoStop}%
\bibitem [{\citenamefont {Hele}\ \emph {et~al.}(2015{\natexlab{b}})\citenamefont {Hele}, \citenamefont {Willatt}, \citenamefont {Muolo},\ and\ \citenamefont {Althorpe}}]{heleCommunicationRelationCentroid2015}%
  \BibitemOpen
  \bibfield  {author} {\bibinfo {author} {\bibfnamefont {T.~J.~H.}\ \bibnamefont {Hele}}, \bibinfo {author} {\bibfnamefont {M.~J.}\ \bibnamefont {Willatt}}, \bibinfo {author} {\bibfnamefont {A.}~\bibnamefont {Muolo}}, \ and\ \bibinfo {author} {\bibfnamefont {S.~C.}\ \bibnamefont {Althorpe}},\ }\href {\doibase 10.1063/1.4921234} {\bibfield  {journal} {\bibinfo  {journal} {J. Chem. Phys.}\ }\textbf {\bibinfo {volume} {142}},\ \bibinfo {pages} {191101} (\bibinfo {year} {2015}{\natexlab{b}})}\BibitemShut {NoStop}%
\bibitem [{\citenamefont {Hele}(2016)}]{Hele2016a}%
  \BibitemOpen
  \bibfield  {author} {\bibinfo {author} {\bibfnamefont {T.~J.~H.}\ \bibnamefont {Hele}},\ }\href {\doibase 10.1080/00268976.2015.1136003} {\bibfield  {journal} {\bibinfo  {journal} {Mol. Phys.}\ }\textbf {\bibinfo {volume} {114}},\ \bibinfo {pages} {1461} (\bibinfo {year} {2016})}\BibitemShut {NoStop}%
\bibitem [{\citenamefont {Jang}\ and\ \citenamefont {Voth}(1999{\natexlab{a}})}]{jang_derivation_1999}%
  \BibitemOpen
  \bibfield  {author} {\bibinfo {author} {\bibfnamefont {S.}~\bibnamefont {Jang}}\ and\ \bibinfo {author} {\bibfnamefont {G.~A.}\ \bibnamefont {Voth}},\ }\href {\doibase 10.1063/1.479515} {\bibfield  {journal} {\bibinfo  {journal} {J. Chem. Phys.}\ }\textbf {\bibinfo {volume} {111}},\ \bibinfo {pages} {2371} (\bibinfo {year} {1999}{\natexlab{a}})}\BibitemShut {NoStop}%
\bibitem [{\citenamefont {Jang}\ and\ \citenamefont {Voth}(1999{\natexlab{b}})}]{jang_path_1999}%
  \BibitemOpen
  \bibfield  {author} {\bibinfo {author} {\bibfnamefont {S.}~\bibnamefont {Jang}}\ and\ \bibinfo {author} {\bibfnamefont {G.~A.}\ \bibnamefont {Voth}},\ }\href {\doibase 10.1063/1.479514} {\bibfield  {journal} {\bibinfo  {journal} {J. Chem. Phys.}\ }\textbf {\bibinfo {volume} {111}},\ \bibinfo {pages} {2357} (\bibinfo {year} {1999}{\natexlab{b}})}\BibitemShut {NoStop}%
\bibitem [{\citenamefont {Hele}\ and\ \citenamefont {Suleimanov}(2015)}]{Hele2015c}%
  \BibitemOpen
  \bibfield  {author} {\bibinfo {author} {\bibfnamefont {T.~J.~H.}\ \bibnamefont {Hele}}\ and\ \bibinfo {author} {\bibfnamefont {Y.~V.}\ \bibnamefont {Suleimanov}},\ }\href {\doibase 10.1063/1.4928599} {\bibfield  {journal} {\bibinfo  {journal} {J. Chem. Phys.}\ }\textbf {\bibinfo {volume} {143}},\ \bibinfo {pages} {074107} (\bibinfo {year} {2015})}\BibitemShut {NoStop}%
\bibitem [{\citenamefont {Kubo}(1966)}]{kuboFluctuationdissipationTheorem1966}%
  \BibitemOpen
  \bibfield  {author} {\bibinfo {author} {\bibfnamefont {R.}~\bibnamefont {Kubo}},\ }\href {\doibase 10.1088/0034-4885/29/1/306} {\bibfield  {journal} {\bibinfo  {journal} {Rep. Prog. Phys.}\ }\textbf {\bibinfo {volume} {29}},\ \bibinfo {pages} {255} (\bibinfo {year} {1966})}\BibitemShut {NoStop}%
\bibitem [{\citenamefont {Hele}\ and\ \citenamefont {Ananth}(2016)}]{Hele2016}%
  \BibitemOpen
  \bibfield  {author} {\bibinfo {author} {\bibfnamefont {T.~J.}\ \bibnamefont {Hele}}\ and\ \bibinfo {author} {\bibfnamefont {N.}~\bibnamefont {Ananth}},\ }\href {\doibase 10.1039/c6fd00106h} {\bibfield  {journal} {\bibinfo  {journal} {Faraday Discuss.}\ }\textbf {\bibinfo {volume} {195}},\ \bibinfo {pages} {269} (\bibinfo {year} {2016})}\BibitemShut {NoStop}%
\bibitem [{\citenamefont {Wigner}(1932)}]{Wigner1932}%
  \BibitemOpen
  \bibfield  {author} {\bibinfo {author} {\bibfnamefont {E.}~\bibnamefont {Wigner}},\ }\href {\doibase 10.1103/PhysRev.40.749} {\bibfield  {journal} {\bibinfo  {journal} {Phys. Rev.}\ }\textbf {\bibinfo {volume} {40}},\ \bibinfo {pages} {749} (\bibinfo {year} {1932})}\BibitemShut {NoStop}%
\bibitem [{\citenamefont {Hillery}\ \emph {et~al.}(1984)\citenamefont {Hillery}, \citenamefont {O'Connell}, \citenamefont {Scully},\ and\ \citenamefont {Wigner}}]{hilleryDistributionFunctionsPhysics1984}%
  \BibitemOpen
  \bibfield  {author} {\bibinfo {author} {\bibfnamefont {M.}~\bibnamefont {Hillery}}, \bibinfo {author} {\bibfnamefont {R.}~\bibnamefont {O'Connell}}, \bibinfo {author} {\bibfnamefont {M.}~\bibnamefont {Scully}}, \ and\ \bibinfo {author} {\bibfnamefont {E.}~\bibnamefont {Wigner}},\ }\href {\doibase 10.1016/0370-1573(84)90160-1} {\bibfield  {journal} {\bibinfo  {journal} {Phys. Rep.}\ }\textbf {\bibinfo {volume} {106}},\ \bibinfo {pages} {121} (\bibinfo {year} {1984})}\BibitemShut {NoStop}%
\bibitem [{\citenamefont {Ceperley}(1995)}]{Ceperley1995}%
  \BibitemOpen
  \bibfield  {author} {\bibinfo {author} {\bibfnamefont {D.~M.}\ \bibnamefont {Ceperley}},\ }\href {\doibase 10.1103/RevModPhys.67.279} {\bibfield  {journal} {\bibinfo  {journal} {Rev. Mod. Phys.}\ }\textbf {\bibinfo {volume} {67}},\ \bibinfo {pages} {279} (\bibinfo {year} {1995})}\BibitemShut {NoStop}%
\bibitem [{\citenamefont {Chakravarty}(1997)}]{chakravartyPathIntegralSimulations1997}%
  \BibitemOpen
  \bibfield  {author} {\bibinfo {author} {\bibfnamefont {C.}~\bibnamefont {Chakravarty}},\ }\href {\doibase 10.1080/014423597230190} {\bibfield  {journal} {\bibinfo  {journal} {Int. Rev. Phys. Chem.}\ }\textbf {\bibinfo {volume} {16}},\ \bibinfo {pages} {421} (\bibinfo {year} {1997})}\BibitemShut {NoStop}%
\bibitem [{\citenamefont {Chakravarty}, \citenamefont {Gordillo},\ and\ \citenamefont {Ceperley}(1998)}]{chakravartyComparisonEfficiencyFourier1998}%
  \BibitemOpen
  \bibfield  {author} {\bibinfo {author} {\bibfnamefont {C.}~\bibnamefont {Chakravarty}}, \bibinfo {author} {\bibfnamefont {M.~C.}\ \bibnamefont {Gordillo}}, \ and\ \bibinfo {author} {\bibfnamefont {D.~M.}\ \bibnamefont {Ceperley}},\ }\href {\doibase 10.1063/1.476725} {\bibfield  {journal} {\bibinfo  {journal} {The Journal of Chemical Physics}\ }\textbf {\bibinfo {volume} {109}},\ \bibinfo {pages} {2123} (\bibinfo {year} {1998})}\BibitemShut {NoStop}%
\bibitem [{\citenamefont {Freeman}\ and\ \citenamefont {Doll}(1984)}]{freemanMonteCarloMethod1984}%
  \BibitemOpen
  \bibfield  {author} {\bibinfo {author} {\bibfnamefont {D.~L.}\ \bibnamefont {Freeman}}\ and\ \bibinfo {author} {\bibfnamefont {J.~D.}\ \bibnamefont {Doll}},\ }\href {\doibase 10.1063/1.446640} {\bibfield  {journal} {\bibinfo  {journal} {J. Chem. Phys.}\ }\textbf {\bibinfo {volume} {80}},\ \bibinfo {pages} {5709} (\bibinfo {year} {1984})}\BibitemShut {NoStop}%
\bibitem [{\citenamefont {Goldstein}(1980)}]{goldsteinClassicalMechanics1980}%
  \BibitemOpen
  \bibfield  {author} {\bibinfo {author} {\bibfnamefont {H.}~\bibnamefont {Goldstein}},\ }\href@noop {} {\emph {\bibinfo {title} {Classical {{Mechanics}}}}},\ \bibinfo {edition} {2nd}\ ed.\ (\bibinfo  {publisher} {Addison-Wesley},\ \bibinfo {address} {Reading, Massachusetts},\ \bibinfo {year} {1980})\BibitemShut {NoStop}%
\bibitem [{\citenamefont {Moyal}(1949)}]{Moyal1949}%
  \BibitemOpen
  \bibfield  {author} {\bibinfo {author} {\bibfnamefont {J.~E.}\ \bibnamefont {Moyal}},\ }\href {\doibase 10.1017/S0305004100000487} {\bibfield  {journal} {\bibinfo  {journal} {Math. Proc. Camb. Philos. Soc.}\ }\textbf {\bibinfo {volume} {45}},\ \bibinfo {pages} {99} (\bibinfo {year} {1949})}\BibitemShut {NoStop}%
\bibitem [{\citenamefont {Zwanzig}(2001)}]{Zwanzig2001}%
  \BibitemOpen
  \bibfield  {author} {\bibinfo {author} {\bibfnamefont {R.}~\bibnamefont {Zwanzig}},\ }\href@noop {} {\emph {\bibinfo {title} {Nonequilibrium Statistical Mechanics}}}\ (\bibinfo  {publisher} {Oxford University Press, New York},\ \bibinfo {year} {2001})\BibitemShut {NoStop}%
\bibitem [{\citenamefont {Bossion}, \citenamefont {Chowdhury},\ and\ \citenamefont {Huo}(2021)}]{Bossion2021}%
  \BibitemOpen
  \bibfield  {author} {\bibinfo {author} {\bibfnamefont {D.}~\bibnamefont {Bossion}}, \bibinfo {author} {\bibfnamefont {S.~N.}\ \bibnamefont {Chowdhury}}, \ and\ \bibinfo {author} {\bibfnamefont {P.}~\bibnamefont {Huo}},\ }\href {\doibase 10.1063/5.0051456} {\bibfield  {journal} {\bibinfo  {journal} {J. Chem. Phys.}\ }\textbf {\bibinfo {volume} {154}},\ \bibinfo {pages} {184106} (\bibinfo {year} {2021})}\BibitemShut {NoStop}%
\bibitem [{\citenamefont {Bossion}\ \emph {et~al.}(2022)\citenamefont {Bossion}, \citenamefont {Ying}, \citenamefont {Chowdhury},\ and\ \citenamefont {Huo}}]{bossion_non-adiabatic_2022}%
  \BibitemOpen
  \bibfield  {author} {\bibinfo {author} {\bibfnamefont {D.}~\bibnamefont {Bossion}}, \bibinfo {author} {\bibfnamefont {W.}~\bibnamefont {Ying}}, \bibinfo {author} {\bibfnamefont {S.~N.}\ \bibnamefont {Chowdhury}}, \ and\ \bibinfo {author} {\bibfnamefont {P.}~\bibnamefont {Huo}},\ }\href {\doibase 10.1063/5.0094893} {\bibfield  {journal} {\bibinfo  {journal} {J. Chem. Phys.}\ }\textbf {\bibinfo {volume} {157}},\ \bibinfo {pages} {084105} (\bibinfo {year} {2022})}\BibitemShut {NoStop}%
\bibitem [{\citenamefont {Kelly}\ \emph {et~al.}(2012)\citenamefont {Kelly}, \citenamefont {{van Zon}}, \citenamefont {Schofield},\ and\ \citenamefont {Kapral}}]{Kelly2012}%
  \BibitemOpen
  \bibfield  {author} {\bibinfo {author} {\bibfnamefont {A.}~\bibnamefont {Kelly}}, \bibinfo {author} {\bibfnamefont {R.}~\bibnamefont {{van Zon}}}, \bibinfo {author} {\bibfnamefont {J.}~\bibnamefont {Schofield}}, \ and\ \bibinfo {author} {\bibfnamefont {R.}~\bibnamefont {Kapral}},\ }\href {\doibase 10.1063/1.3685420} {\bibfield  {journal} {\bibinfo  {journal} {J. Chem. Phys.}\ }\textbf {\bibinfo {volume} {136}},\ \bibinfo {pages} {084101} (\bibinfo {year} {2012})}\BibitemShut {NoStop}%
\bibitem [{\citenamefont {Saller}, \citenamefont {Kelly},\ and\ \citenamefont {Richardson}(2020)}]{saller_improved_2020}%
  \BibitemOpen
  \bibfield  {author} {\bibinfo {author} {\bibfnamefont {M.~A.~C.}\ \bibnamefont {Saller}}, \bibinfo {author} {\bibfnamefont {A.}~\bibnamefont {Kelly}}, \ and\ \bibinfo {author} {\bibfnamefont {J.~O.}\ \bibnamefont {Richardson}},\ }\href {\doibase 10.1039/C9FD00050J} {\bibfield  {journal} {\bibinfo  {journal} {Faraday Discuss.}\ }\textbf {\bibinfo {volume} {221}},\ \bibinfo {pages} {150} (\bibinfo {year} {2020})}\BibitemShut {NoStop}%
\bibitem [{\citenamefont {Huo}\ and\ \citenamefont {Coker}(2011)}]{Huo2011}%
  \BibitemOpen
  \bibfield  {author} {\bibinfo {author} {\bibfnamefont {P.}~\bibnamefont {Huo}}\ and\ \bibinfo {author} {\bibfnamefont {D.~F.}\ \bibnamefont {Coker}},\ }\href {\doibase 10.1063/1.3664763} {\bibfield  {journal} {\bibinfo  {journal} {J. Chem. Phys.}\ }\textbf {\bibinfo {volume} {135}},\ \bibinfo {pages} {201101} (\bibinfo {year} {2011})}\BibitemShut {NoStop}%
\bibitem [{\citenamefont {Kirrander}\ and\ \citenamefont {Vacher}(2020)}]{kirrander_ehrenfest_2020}%
  \BibitemOpen
  \bibfield  {author} {\bibinfo {author} {\bibfnamefont {A.}~\bibnamefont {Kirrander}}\ and\ \bibinfo {author} {\bibfnamefont {M.}~\bibnamefont {Vacher}},\ }in\ \href {\doibase 10.1002/9781119417774.ch15} {\emph {\bibinfo {booktitle} {Quantum {Chemistry} and {Dynamics} of {Excited} {States}}}}\ (\bibinfo  {publisher} {John Wiley \& Sons, Ltd},\ \bibinfo {year} {2020})\ pp.\ \bibinfo {pages} {469--497}\BibitemShut {NoStop}%
\end{thebibliography}%


\end{document}